\documentclass{llncs}
\usepackage[utf8]{inputenc}

\usepackage[nocompress]{cite}
\usepackage{lmodern}
\usepackage{amsmath,amssymb,stmaryrd}
\usepackage{wrapfig}
\usepackage[usenames,dvipsnames,pdftex]{color}
\usepackage[colorlinks,pagebackref,linkcolor=blue,citecolor=BrickRed,urlcolor=OliveGreen]{hyperref}

\usepackage[center,noinfo, cross, width=165mm, height=220mm]{crop}

\usepackage[T1]{fontenc}
\usepackage{calc,pifont}
\usepackage{xcolor}
\usepackage{graphicx}
\usepackage{fancyhdr}
\usepackage{array}
\newcolumntype{L}[1]{m{#1}}

\usepackage{multirow}
\usepackage{longtable}
\usepackage{algorithm}
\usepackage[noend]{algpseudocode}
\usepackage{tikz}
\usepackage{xspace}
\usepackage{caption}
\usepackage{compactspacing}

\usepackage{etoolbox}
\newtoggle{longv}

\toggletrue{longv}

\newcommand\longv[2][]{\iftoggle{longv}{#2}{#1}}

\usepackage{appendix}
\capturecounter{theorem}
\capturecounter{figure}

\usepackage{todonotes}

\long\def\ignore#1{\relax}

\newcommand\struto[1][15pt]{{\raise #1 \hbox{\strut}}}%
\newcommand\strutb[1][15pt]{{\raise-#1 \hbox{\strut}}}%

\makeatletter
\def\@boxfigurewith[#1]{\figure[#1]\vbox\bgroup\hrule height.1em}
\def\@boxfigurewithout{\figure\vbox\bgroup\hrule height.1em}

\makeatother

\renewcommand\[[1][3]{\par\removelastskip\vskip#1pt\vbox\bgroup\hrule height0pt\vfil\hbox to\hsize\bgroup\hfil\(}
\renewcommand\][1][3]{\)\hfil\egroup\vfil\hrule height0pt\egroup\vskip#1pt\nointerlineskip\noindent}

\newbox\columnsbox
\newbox\tmpbox
\newdimen\columnsheight
\newdimen\columnwidth
\newdimen\remainingwidth
\newdimen\textwidthsave
\def\mycolumnsheight{}

\newcommand\columns[1]{%
  \def\mycolumnsheight{}%
  \setlength\remainingwidth\textwidth%
  \setbox\columnsbox=\vbox\bgroup\vskip0pt\vfil\hbox to\textwidth\bgroup#1\egroup\vfil\egroup%
  \columnsheight=\ht\columnsbox%
  \def\mycolumnsheight{to\columnsheight}%
  \hrule height 0pt\vtop{\hbox to\wd\columnsbox\bgroup#1\egroup}%
}

\makeatletter

\def\commonpart{%
  \setlength\columnwidth{\wd\tmpbox}%
  \vtop{\vskip0pt\hbox to\columnwidth{{\box\tmpbox}}}%
  \advance\remainingwidth-\columnwidth%
  \setlength\textwidth\textwidthsave%
  \hsize\textwidthsave%
}
\def\column{\unskip\setlength\textwidthsave\textwidth\@ifnextchar[\@columnwith\@columnwithout}
\long\def\@columnwith[#1]#2{%
  \def\newhsize{#1\dimexpr\textwidth\relax}%
  \hsize\newhsize%
  \ifdim\hsize<0.1pt\hsize\remainingwidth\fi%
  \setlength\textwidth\hsize%
  \setbox\tmpbox=\hbox to\hsize\bgroup\hfil\vtop\mycolumnsheight{\vskip0pt#2\vskip0pt}\hfil\egroup%
  \commonpart%
}
\long\def\@columnwithout#1{%
  \hsize\remainingwidth%
  \setlength\textwidth\hsize%
  \setbox\tmpbox=\hbox\bgroup\vtop\mycolumnsheight{\vskip0pt#1\vskip0pt}\egroup%
  \commonpart%
}
\makeatother

\newcommand\abs[1]{{\left|#1\right|}}        % val.absolue
\newcommand{\eqdef}{:= }
\newcommand{\recdef}{::=\ }

\newcommand{\sep}{\mbox{\,\,\,\rule[-.35\ht\strutbox]{.7pt}{1.3\ht\strutbox}\,\,\,}}    % Separators in BNF definitions

\newcommand\CDCL{\textsf{CDCL}\xspace}
\newcommand\CDCLT{\textsf{CDCL}$(\mathcal T)$\xspace}
\newcommand\MCSAT{\textsf{MCSAT}\xspace}

\renewcommand{\iff}{if and only if}

\newcommand{\ie}{i.e.,\xspace}

\newcommand{\eg}{e.g.,\xspace}

\newcommand{\resp}{resp.\xspace}

\let\oldtimes\times
\renewcommand\times{{\oldtimes}}
\renewcommand\prod{{\cdot}}

\newcommand\col{\!:\!}

\newcommand\eq[1][]{\ensuremath{{\,\simeq_{#1}\,}}}
\newcommand\noteq[1][]{\ensuremath{{\,\not\simeq_{#1}\,}}}

\newcommand\imp{\Rightarrow}
\newcommand\equivalent{\Leftrightarrow}

\makeatletter
\newcommand\fv[2][]{
  \def\@fvwith[##1]{{\sf var}_{##1}^{#1}(#2)}
  \def\@fvwithout{{\sf var}^{#1}(#2)}
  \@ifnextchar[\@fvwith\@fvwithout
  }
\makeatother

\newcommand\sem[2]{[\![{#2}]\!]_{{#1}}}
\makeatletter
\def\extract#1#2{%
  \def\@extractOne{#1[#2]}%
  \def\@extractTopLow[##1]{#1[#2{\,\col\,}##1]}%
  \@ifnextchar[\@extractTopLow\@extractOne%
  }
\makeatother

\newcommand\indexset[1]{\mathfrak{#1}}
\newcommand\sliced[1]{{#1}_{s}}
\newcommand\evaluates[2][]{y\not\in\fv{#2}}
\newcommand\eval[2][]{\sem{#1}{#2}}

\renewcommand\implies{\Rightarrow}

\newcommand\vtrue{{\sf true}}

\newcommand\ZZ{\mathbb Z}

\newcommand\BVth[1][]{\ensuremath{\mathcal B\mathcal V_{#1}}\xspace}

\newcommand\uleq{\leq^{\sf u}}
\newcommand\sleq{\leq^{\sf s}}
\newcommand\ult{<^{\sf u}}
\newcommand\slt{<^{\sf s}}

\makeatletter
\newcommand\decor{
  \def\@decorwith[##1]##2{{}_{##1{\vdash}}{##2}}
  \def\@decorwithout##1{{}_{?}{##1}}
  \@ifnextchar[\@decorwith\@decorwithout
  }
\makeatother

\newcommand\interval[2]{{\textbf [}#1\,{;}\,#2{\textbf [}}
\newcommand\full[1][]{{\sf full}^{#1}}
\newcommand\empt[1][]{\interval 0 0}

\newcommand\sextension[2]{\mbox{$\pm$-ext}_{#1}(#2)}
\newcommand\bvnot[1]{\mbox{bvnot}(#1)}

\makeatletter
\def\extract#1#2{%
  \def\@extractOne{#1[#2]}%
  \def\@extractTopLow[##1]{#1[#2{\,\col\,}##1]}%
  \@ifnextchar[\@extractTopLow\@extractOne%
  }
\makeatother

\newcommand\bextract[2]{\extract{#2}{#1}[]}
\newcommand\uextract[2]{\extract{#2}{}[#1]}

\newcommand\bwidth[1]{\abs{#1}}
\newcommand\Iinversename{\textsf{forbid}}
\newcommand\Iinverse[3]{\Iinversename(\,#1\,,\,#2\,,\,#3\,)}
\newcommand\Iuptrim[4]{\textsf{utrim}_{#1}(\,#2\,,\,#3\,,\,#4\,)}
\newcommand\Idowntrim[4]{\textsf{dtrim}_{#1}(\,#2\,,\,#3\,,\,#4\,)}

\title{Solving bitvectors with MCSAT:\\ explanations from bits and pieces\longv{ (long version)}}
\titlerunning{Solving bitvectors with MCSAT: explanations from bits and pieces\longv{ (long version)}}

\author{
  Stéphane Graham-Lengrand
  \and Dejan Jovanović
  \and Bruno Dutertre
}

\institute{
  SRI International, USA
}

\authorrunning{Stéphane Graham-Lengrand, Dejan Jovanović, and Bruno Dutertre}

\begin{document}

\maketitle

\begin{abstract}
  We present a decision procedure for the theory of fixed-sized bitvectors in
  the MCSAT framework. MCSAT is an alternative to CDCL(T) for SMT solving and
  can be seen as an extension of CDCL to domains other than the Booleans. Our
  procedure uses BDDs to record and update the sets of feasible values of
  bitvector variables. For explaining conflicts and propagations, we develop
  specialized word-level interpolation for two common fragments of the theory.
  For full generality, explaining conflicts outside of the covered fragments resorts to local
  bitblasting. The approach is implemented in the
  Yices 2 SMT solver and we present experimental results.
\end{abstract}

\thispagestyle{fancy}

% !TEX root = main.tex
% !TEX program = pdflatex

\section{Introduction}

\emph{Model-constructing satisfiability}
(\MCSAT)~\cite{Jovanovic13mcsat,JBdM13,Jovanovic:vmcai17} is an alternative to
the \CDCLT scheme \cite{Nieuwenhuis06} for Satisfiability Modulo Theories (SMT).
While \CDCLT interfaces a \CDCL SAT solver~\cite{MSLM:HS:2009} with black-box
decision procedures, \MCSAT integrates first-order reasoning into \CDCL
directly. Like \CDCL, \MCSAT alternates between search and conflict analysis.
In the search phase, \MCSAT assigns values to first-order variables and
propagates unit consequences of these assignments. If a conflict occurs during
search, \eg when the domain of a first-order variable is empty,
\MCSAT enters conflict analysis and learns an explanation, which is a
symbolic representation of what was wrong with the assignments causing the
conflict. As in \CDCL, the learned clause triggers backtracking from which
search can resume. Decision procedures based on \MCSAT have demonstrated
strong performance in theories such as non-linear real \cite{Jovanovic13mcsat} and
integer arithmetic \cite{Jovanovic:vmcai17}. These theories are relatively
well-behaved and provide features such as quantifier elimination and
interpolation---the building blocks of conflict resolution in \MCSAT.

We describe an \MCSAT decision procedure for the theory of bitvectors
(\BVth). In contrast to arithmetic, the complexity of \BVth in terms
of syntax and semantics, combined with the lack of word-level
interpolation and quantifier elimination, makes the development of
\BVth decision procedures (\MCSAT or not) very difficult. The
state-of-the art \BVth decision procedures are all based on a
``preprocess and bitblast'' pipeline
\cite{ganesh2007decision,niemetz2014boolector,kroening2016decision}:
they reduce the \BVth problems to a pure SAT problem by reducing the
word-level semantics to bit-level semantics. Exceptions to the
bitblasting approach do exist,
such as~\cite{Bruttomesso:ICCAD09,hadarean2014tale},
which generally do not perform as well as bitblasting
except on small classes of crafted examples,
and the \MCSAT approach of~\cite{ZeljicWR16},
which we discuss below and in the conclusion.

An \MCSAT decision procedure must provide two theory-specific reasoning
mechanisms.

%which left open the idea of 

First, the procedure must maintain a set of values that are feasible
for each variable. This set is updated during the search. It is used
to propagate variable values and to detect a conflict when the set
becomes empty. Finding a suitable representation for domains is a key
step in integrating a theory into \MCSAT. We represent variable
domains with Binary Decision Diagrams
(BDDs)~\cite{bryant1986graph}. BDDs can represent any set of bitvector
values. By being canonical, they offer a simple mechanism to detect
when a domain becomes a singleton---in which case \MCSAT can perform a
theory propagation---and when a domain becomes empty--in which case
\MCSAT enters conflict analysis. In short, BDDs offer a generic
mechanism for proposing and propagating values, and for detecting
conflicts. In contrast, previous work by Zelji\'{c} et
al.~\cite{ZeljicWR16} represents bitvector domains using
\emph{intervals} and \emph{patterns}, which cannot represent every set
of bitvector values precisely; they over-approximate the domains.

Second, once a conflict has been detected, the procedure must construct a
symbolic explanation of the conflict. This explanation must rule out
the partial assignment that caused the conflict, but it is
desirable for explanations to generalize and rule out larger parts of
the search space. For this purpose, previous work~\cite{ZeljicWR16}
relied on incomplete abstraction techniques (replace a value by an
interval; extend a value into a larger set by leaving some bits
unassigned),
and left open the idea of using interpolation to produce explanations.
%% Knowing that we cannot possible provide a uniform
%% word-level explanation mechanism,
Instead of aiming for a uniform, generic explanation mechanism, we
take a modular approach.  We develop efficient word-level explanation
procedures for two useful fragments of \BVth, based on interpolation.
%% explanations can be built. and apply them to conflicts that fit into
%% these fragments.%
%
% \footnote{Note that our notion of `fit` considers model-based normalizations and
% is more general than syntax matching.\todo[inline]{add example for `fit`}}
%
Our first fragment includes bitvector
equalities, extractions, and concatenations where word-level explanations can be
constructed through model-based variants of classic equality reasoning
techniques (e.g.,~\cite{Cyrluk1997,Bruttomesso:ICCAD09,detlefs2005simplify}).
Our second fragment is a subset of linear arithmetic where explanations are
constructed by interval reasoning in modular arithmetic. When conflicts
do not fit into either fragment, we build an explanation by
% our baseline explanation mechanism is to
bitblasting and extracting an unsat core.
Although this fallback produces theory lemmas expressed
at the bit-level, it is used only as a last resort. In addition, this
bitblasting-based procedure is local and limited to constraints that are relevant
to the current conflict; we do not apply bitblasting to the full problem.

%% this last resort bitblasting is only used for constraints that
%% do not live in covered fragments of $\BVth$. In addition, the bitblasting is
%% only applied to the local conflict and not to the overall problem.

Section~\ref{sec:generalscheme}, is an overview of \MCSAT. It also presents the
BDD approach and general considerations for conflict explanation.
Section~\ref{sec:concatextract} describes our interpolation algorithm for
equality with concatenation and extraction.  Section~\ref{sec:bvarith} presents
our interpolation method for a fragment of linear bitvector arithmetic.
Section~\ref{sec:normalisation} presents the normalization technique
we apply to conflicts in the hope of expressing them in that bitvector arithmetic fragment.
Section~\ref{sec::experiments} presents an evaluation
of the approach, which we implemented in the Yices~2 solver~\cite{Dutertre14yices}.\footnote{This paper extends
preliminary results presented at the SMT
workshop~\cite{GrahamLengrandJovanovicSMT17,GrahamLengrandJovanovicSMT19} and
includes a full implementation and experimental evaluation.}

% !TEX root = main.tex
% !TEX program = pdflatex

\section{A General Scheme for Bitvectors}
\label{sec:generalscheme}

By $\BVth$, we denote the theory of quantifier-free fixed-sized
bitvectors, a.k.a.\ QF\_BV in SMT-LIB~\cite{BarST-SMTLIB}.
A first-order term $u$ of $\BVth$ is sorted as either a Boolean
or a bitvector of a fixed \emph{length} (a.k.a.~\emph{bitwidth}), denoted $\bwidth u$.
Its set of variables (a.k.a.~uninterpreted constants) is denoted $\fv u$.
This paper only uses a few $\BVth$ operators.
The concatenation of bitvector terms $t$ and $u$ is denoted $t\circ u$;
the binary predicates $\ult$, $\uleq$ denote unsigned comparisons, and
$\slt$, $\sleq$ denote signed comparisons.
In such comparisons, both operands must have the same bitwidth.
If $n$ is the bitwidth of $u$, and $l$ and $h$ are two integer indices such
that $0 \leq l < h \leq n$, then $\extract u h [l]$, extracts $h{-}l$
bits of $u$, namely the bits at indices between $l$ and $h{-}1$
(included). We write $\uextract l u$, $\bextract h u$, and
$\extract u l$
as abbreviations for $\extract u n [l]$, $\extract u h [0]$, and $\extract u {l{+}1}[l]$,
respectively.
Our convention is to have bitvector indices start from the right-hand side,
so that bit $0$ is the right-most bit and $\bextract 2 {0011}$ is $11$.
We use standard notations for bitvector arithmetic, which coincides with
arithmetic modulo $2^w$ where $w$ is the bitwidth.  We sometimes use integer
constants \eg $0$, $1$, $-1$ for bitvectors when the bitwidth is clear.
We use the standard (quantifier-free) notions of
\emph{literal}, \emph{clause}, \emph{cube}, and \emph{formula}~\cite{RobinsonV01}.

A \emph{model} of
a $\BVth$ formula $\Phi$ is an assignment that gives a bitvector (\resp Boolean) value
to all bitvector (\resp Boolean) variables of $\Phi$,
in such a way that $\Phi$ evaluates to true,
under the standard interpretation of Boolean and bitvector symbols.
To simplify the presentation, we assume in this paper that there are no Boolean variables,
although they are supported in our implementation.

\subsection{MCSAT Overview}

\newcommand{\dtrail}[1]{\llbracket\; #1 \;\rrbracket}
\newcommand{\ddec}[2]{\ensuremath{{#1} \mapsto #2}}
\newcommand{\dprop}[3]{\ensuremath{#1 \overset{#3}{\leadsto} #2}}

%% \MCSAT takes as input a quantifier-free first-order formula and determines whether
%% the formula is satisfiable in a given theory.
%% In the present case of theory $\BVth$,
%% it is satisfiable if it has a \emph{model},
%% \ie an assignment of bitvector (\resp Boolean) values to
%% free bitvector (\resp Boolean) variables,
%% so that the input formula evaluates to true (using the %standard model and
%% standard interpretation of Boolean and bitvector symbols).

%% Building on the \CDCL ideas~\cite{MSLM:HS:2009},

\MCSAT\ searches for a model of an input quantifier-free formula by building a partial
assignment---maintained in a \emph{trail}---and extends the concepts of unit
propagation and consistency to first-order terms and
literals~\cite{Jovanovic:vmcai17,Jovanovic13mcsat,JBdM13}.
Reasoning is implemented by theory-specific plugins, each of which has
a partial view of the trail. In the case of $\BVth$, the bitvector
plugin sees in the trail an assignment $\mathcal{M}$ of the
form $x_1\mapsto v_1,\ldots,x_n\mapsto v_n$ that gives values to
bitvector variables, and a set of bitvector literals $L_1,\ldots,L_t$,
called \emph{constraints}, that must be true in the current trail.
\MCSAT\ and its bitvector plugin
maintain the invariant that none of the literals $L_i$
evaluates to false under $\mathcal{M}$; either $L_i$ is true or some
variable of $L_i$ has no value in $\mathcal{M}$.
To maintain this invariant, they detect \emph{unit inconsistencies\/}:
We say that literal $L_i$ is \emph{unit in $y$} if $y$ is the only unassigned variable of $L_i$,
and that a trail is \emph{unit inconsistent\/} if there is a variable $y$
and a subset $\{ C_1,\ldots, C_m\}$ of $\{ L_1,\ldots, L_t \}$,
called a \emph{conflict}, such that
every $C_j$ is unit in $y$ and the formula $\exists y \bigwedge_{i=1}^mC_i$
evaluates to false under $\mathcal{M}$.  In such a
case, $y$ is called the \emph{conflict variable\/} and $C_1,\ldots,C_m$
are called the \emph{conflict literals\/}.

When such a conflict is detected, the current assignment, or partial model, $\mathcal{M}$
cannot be extended to a full model; some values assigned to
$x_1,\ldots, x_n$ must be revised. As in \CDCL, \MCSAT backtracks and
updates the current assignment by learning a new clause that explains
the conflict.  This new clause must not contain other variables than
$x_1,\ldots,x_n$ and it must rule out the current assignment. For some
theories, this \emph{conflict explanation\/} can be built by
quantifier elimination. More generally, we can build an explanation from
an \emph{interpolant}.
\begin{definition}[Interpolant]\label{def:interpolation}
  A clause $I$ is an \emph{interpolant}\footnote{This is
    the same as the usual notion of (reverse) interpolant between formulas if
    we see $\mathcal{M}$ as the formula $F_{\mathcal M}$ defined
    by $(x_1 \eq v_1) \wedge \cdots \wedge (x_n \eq v_n)$: the
    interpolant is implied by $F$, it is inconsistent with
    $F_{\mathcal M}$, and its variables occur in both $F$ and
    $F_{\mathcal M}$.}
  for formula $F$ at model $\mathcal{M}$ assigning values to $x_1,\ldots, x_n$, if
  (1) $F \Rightarrow I$ is valid (in \BVth),
  (2) The variables in $I$ are in $\{x_1,\ldots, x_n\}\cap \fv F$, and
  (3) $I$ evaluates to false in $\mathcal{M}$.
\end{definition}
Given an interpolant $I$ for the conjunction $\bigwedge_{i=1}^mC_i$ of the conflict literals
(or equivalently, for $\exists y\bigwedge_{i=1}^mC_i$) at the current model $\mathcal M$,
the conflict explanation is clause
\mbox{$\left(\bigwedge_{i=1}^mC_i\right) \Rightarrow I$}.
Our main goal is constructing such interpolants in \BVth.

\subsection{BDD Representation and Conflict Detection}\label{sec:bdd}

To detect conflicts, we must keep track of the set of feasible values
for every unassigned variable $y$. These sets are frequently updated
during search so an efficient representation is critical. The
following operations are needed:
\begin{itemize}
\item updating the set when a new constraint becomes unit in $y$,
\item detecting when the set becomes empty,
\item selecting a value from the set.
\end{itemize}

For \BVth, Zelji\'{c} et al.~\cite{ZeljicWR16} represent sets of
feasible values using both intervals and bit patterns. For example,
the set defined by the interval $[0000,0011]$ and the pattern $???1$
is the pair $\{ 0001, 0011 \}$ (i.e., all bitvectors in the interval
whose low-order bit is $1$). This representation is lightweight and
efficient but it is not precise. Some sets are not representable
exactly.  We use Binary Decision Diagrams (BDD)~\cite{bryant1986graph}
over the bits of $y$.
The major advantage is that BDDs provide an exact implementation of any set of values for $y$.
Updating sets of values amounts to computing the conjunction of BDDs
(i.e., set intersection).
Checking whether a set is empty and selecting a value in the set (if it is not),
can be done efficiently by, respectively,
checking whether the BDD is false, and performing a
top-down traversal of the BDD data structure.
There is a risk that the BDD
representation explodes but this risk is reduced in our context since
each BDD we build is for a single variable (and most variables do not have too many bits).
We use the CUDD package~\cite{CUDD} to implement BDDs.

\subsection{Baseline Conflict Explanation}

Given a conflict as described previously,
the clause $(x_1 \noteq v_1)\vee \cdots \vee (x_n \noteq v_n)$,
which is falsified by model $\mathcal M$ only,
is an interpolant for $\bigwedge_{i=1}^mC_i$ at $\mathcal{M}$
according to Definition~\ref{def:interpolation}.
This gives the following trivial conflict explanation:
\[
C_1 \wedge \cdots \wedge C_m \implies
(x_1 \noteq v_1) \vee \cdots \vee (x_n \noteq v_n)
\]
We seek to generalize model ${\mathcal M}$ with a formula
that rules out bigger parts of the search space than just ${\mathcal M}$.
A first improvement is replacing the constraints by a \emph{core} $\mathcal{C}$,
that is, a minimal subset of $\{C_1,\ldots,C_n\}$ that evaluates to
false in ${\mathcal M}$.\footnote{In our implementation, we construct
  $\cal{C}$ using the QuickXplain algorithm~\cite{junker2001quickxplain}.}

%% In Yices, we produce conflict clauses as follows:
%% First, we use BDDs to isolate the core $\mathcal{C}$ of the constraints in conflict
%% (relying on the quick-explain mechanism~\cite{junker2001quickxplain}),
%% and then we decide which interpolation procedure to apply
%% depending on the fragment of $\BVth$ where the core $\mathcal{C}$ lives.
%% The smaller the $\mathcal{C}$,
%% the higher the chances that it lives in an isolated fragment of $\BVth$.
%% In Sections~\ref{sec:concatextract} and~\ref{sec:bvarith},
%% we present two interpolation procedures specialized to two fragments of \BVth.

To produce the interpolant $I$, we can bitblast the constraints $C_1,
\ldots, C_m$ and solve the resulting SAT problem \emph{under the
  assumptions} that each bit of $x_1, \ldots, x_n$ is true or false as
indicated by the values $v_1, \ldots, v_n$. Since the SAT problem encodes a conflict,
the SAT solver will return an \emph{unsat core},
from which we can extract bits of $v_1, \ldots, v_n$ that contribute
to unsatisfiability.  This generalizes ${\mathcal M}$ by leaving some bits
unassigned, as in~\cite{ZeljicWR16}.

This method is general. It works whatever the constraints
$C_1,\ldots,C_m$, so we use it as a default procedure. The bitblasting
step focuses on constraints that are unit in $y$, which typically
leads to a much smaller SAT problem than bitblasting the whole
problem from the start. However, the bitblasting approach can still
be costly and it may produce weak explanations.

%% bitblasting always applies, and we use it as the default procedure
%% to produce a conflict clause. Note that we only need to bitblast the
%% set of constraints that were unit in $y$. This can be significantly
%% smaller than the whole set of bitvectors constraints originating from the input formula.

%% Although always applicable,
%% the bitblasting approach is not too appealing as a way of
%% producing explanation. In some cases, when the constraints in
%% $\mathcal{C}$ live in a suitable fragment of $\BVth$, there are better
%% ways to produce explanations. By better we mean that the cost of
%% generating the explanation may be cheaper, and/or the explanation
%% itself may rule out more values.

\begin{example}
Consider the constraints
$\lbrace
    x_1 \noteq x_2, \enspace
    x_1 \eq y, \enspace
    x_2 \eq y
  \rbrace$
and the assignment $x_1 \mapsto 1001, x_2 \mapsto 0101$.
The bitblasting approach might produce explanation
$ (x_1 \eq y \wedge x_2 \eq y) \Rightarrow (\extract{x_1}{3} \Rightarrow \extract{x_2}{3})$.
After backtracking, we might similarly learn that $(\extract{x_2}{3}
\Rightarrow \extract{x_1}{3})$.  In this way, it will take eight
iterations to learn enough information to represent the high-level
explanation:
\[
  (x_1 \eq y \wedge x_2 \eq y) \Rightarrow x_1\eq x_2\enspace.
\]
A procedure that can produce $(x_1\eq x_2)$ directly is much more efficient.
\end{example}

% !TEX root = main.tex
% !TEX program = pdflatex

\section{Equality, Concatenation, Extraction}
\label{sec:concatextract}

Our first specialized interpolation mechanism applies when constraints
$\mathcal{C} = \lbrace C_1,\ldots, C_m \rbrace$ belong to the following grammar:
\[
\begin{array}{l@{\quad}ll}
  \mbox{Constraints} &   C & \recdef t\eq t \mid t\noteq t\\
  \mbox{Terms} & t & \recdef e \mid \extract y h [l] \mid t\circ t
\end{array}
\]
where $e$ ranges over any bitvector terms such that $y\not\in\fv
e$.  Without loss of generality, we can assume that $\mathcal{C}$ is a
core.  We split $\mathcal{C}$ into a set of equalities $E=\{a_i\eq
b_i\}_{i\in\indexset E}$ and a set of disequalities $D=\{a_i\noteq
b_i\}_{i\in\indexset D}$.
%
%% Since $\mathcal{C}$ is a conflict, we know that there is no extension of
%% $\mathcal{M}$ with a value for $y$ that can satisfy it. Furthermore, we can
%% assume that this $\mathcal{C}$ is a core, \ie that no constraint can be removed
%% from $\mathcal{C}$ while remaining a conflict.

\paragraph{Slicing.}
Our first step rewrites $\mathcal{C}$ into an equivalent
\emph{sliced\/} form. This computes the \emph{coarsest-base
  slicing}~\cite{Cyrluk1997,Bruttomesso:ICCAD09} of equalities and
disequalities in $\mathcal{C}$. The goal of this rewriting step is to
split the variables into slices that can be treated as independent
terms. The terms in coarsest-base slicing are either of the form
$\extract{y}h[l]$ (slices), or are \emph{evaluable terms} $e$ with
$y\not\in\fv e$.
\begin{example}
Consider the constraints $E = \lbrace \extract {x_1} 4[0] \eq \extract {x_1} 8[4], \extract y 6[2] \eq \extract y 4[0] \rbrace$ and $\lbrace \extract y 4[0] \noteq \extract{x_1}8[4] \rbrace$ over variables $y$ of length 6, and $x_1$ of length 8.
We cannot treat $\extract y 6[2]$ and $\extract y 4[0]$ as independent terms because they overlap.
To break the overlap, we introduce slices: $\extract y 6[4]$, $\extract y 4[2]$, and $\extract y 2[0]$. Equality $\extract y 6[2] \eq \extract y 4[0]$ is rewritten to
$(\extract y 6[4] \eq \extract y 4[2]) \wedge (\extract y 4[2] \eq \extract y 2[0])$. Disequality $\extract y 4[0] \noteq \extract{x_1}8[4]$ is rewritten to
$(\extract y 4[2] \noteq \extract{x_1}8[6])\vee (\extract y 2[0] \noteq \extract{x_1}6[4])$. The final result is
%
% Note that the slices of $y$ are overlapping. This problem can be transformed though further slicing into a coarsest-base slicing
%
\[
\begin{array}{c}
\sliced E =
\{\
  \extract {x_1} 4[2] \eq \extract {x_1} 8[6]\ ,\
  \extract {x_1} 2[0]  \eq \extract {x_1} 6[4]\ ,\
  \extract y 6[4] \eq \extract y 4[2]\ ,\
  \extract y 4[2] \eq \extract y 2[0]
\ \}\enspace, \\
\sliced D =
\{\
(\extract y 4[2] \noteq \extract{x_1}8[6])\vee
(\extract y 2[0] \noteq \extract{x_1}6[4])
\ \}.
\end{array}
\]
%% \todo[inline]{Not sure why there are constraints that don't contain $y$ in this example. They cannot be in the conflict and it seems irrelevant to slice them.}
%
%% After slicing, the set of equalities $E$ becomes a set of equalities $\sliced
%% E$, while the set of disequalities $D$ becomes a set of clauses $D_s$ containing
%% disequalities.
%
\end{example}

\paragraph{Explanations.}
After slicing, we obtain a set $\sliced E$ of equalities and a set $\sliced D$ that contains disjunctions of disequalities.
We can treat each slice as a separate variable, so the problem lies within the theory of equality on a \emph{finite domain}.
%% The cardinality of the bitvector domain can contribute to the conflict if there are not enough bitvectors to satisfy all disequalities.
\begin{algorithm}[!t]
  \caption{E-graph with value management}\label{algo:egraph}
  \begin{algorithmic}[1]
    \Function{e\_graph}{$\sliced E,\mathcal M$}
    \State\Call{Initialize}{$\mathcal G$}
    \Comment{each evaluable term or slice is its own component}
    \For{$t_1\eq t_2\in \sliced E$}
    \State $t_1'\gets \Call{rep}{t_1,\mathcal G}$
    \Comment{get representative for $t_1$'s component}
    \State $t_2'\gets \Call{rep}{t_2,\mathcal G}$
    \Comment{get representative for $t_2$'s component}
    \If{$\evaluates[\mathcal M]{t_1'}$
      and $\evaluates[\mathcal M]{t_2'}$
      and $\eval[\mathcal M]{t_1'}\neq\eval[\mathcal M]{t_2'}$}
    \State{${\sf raise\_conflict}(E\implies t_1'\eq t_2')$}\Comment{$D$ must be empty}
    \EndIf
    \State $t_3\gets \Call{select}{t_1',t_2'}$
    \Comment{select representative for merged component}
    \State $\mathcal G\gets\Call{merge}{t_1,t_2,t_3,\mathcal G}$
    \Comment{merge the components with representative $t_3$}
    \EndFor
    \State\Return{$\mathcal G$}
    \EndFunction
  \end{algorithmic}
\end{algorithm}

We first analyze the conflict with equality reasoning against the
model, as shown in Algorithm~\ref{algo:egraph}. We construct the
E-graph $\mathcal G$ from $\sliced E$~\cite{detlefs2005simplify},
while also taking into account the partial model $\mathcal{M}$ that
triggered the conflict. The model can evaluate terms $e$ such that
$y\not\in\fv e$ to values $\eval[\mathcal M]{e}$, and those can be the
source of the conflict.
To use the model for evaluating terms, we maintain two invariants during E-graph construction:
\begin{enumerate}
  \item If a component contains an evaluable term $c$, then the representative
  of that component is evaluable.
  \item Two evaluable terms $c_1$ and $c_2$ in the same component must evaluate to the same value, otherwise this is the source of the conflict.
\end{enumerate}
The E-graph construction can detect and explain basic conflicts between the
equalities in $E$ and the current assignment.
\begin{example}
Let $r_1$, $r_2$, $r_3$ be bit ranges of the same width. Let $E$ be such that
$\sliced E = \lbrace \extract{x_1}{r_1}\eq\extract{y}{r_3},\enspace
\extract{x_2}{r_2}\eq\extract{y}{r_3} \rbrace$,  and let $D = \emptyset$.
Consider the model $\mathcal M\eqdef x_1 \mapsto 0\ldots0, x_2 \mapsto
1\ldots1$. Then, \Call{e\_graph}{$\sliced E,\mathcal M$} produces the conflict
clause $E\implies\extract{x_1}{r_1}\eq\extract{x_2}{r_2}$.
\end{example}

If the E-graph construction does not raise a conflict, then $\mathcal
M$ is compatible with the equalities in $\sliced E$.  Since $\mathcal{C}$ conflicts with $\mathcal M$,
the conflict explanation must involve $\sliced D$.
%% The assignemnt $\mathcal M$
%% provides a value to each E-graph component that contains an evaluable
%% term.  Moreover, we can extend $\mathcal M$ into a model of $\sliced
%% E$ by assigning an arbitrary value to other components. Since $\mathcal{C}$ conflicts with $\mathcal M$,
%% any such extension must violate a disjunction in $\sliced D$.
%% %
%% Moreover, $\sliced E$ can be satisfied by giving an arbitrary value to every other
%% component. But, since $\mathcal{C}$ is a conflict, any assignment of values to
%% components that don't evaluate must be inconsistent with the clauses in $\sliced
%% D$, \ie the conflict involves the disequalities.
To obtain an explanation, we decompose each disjunct
$C \in \sliced D$ into $\left( C_{\sliced E}\vee C_{\mathcal M}\vee C_{\sf interface}\vee C_{\sf free} \right)$ as follows.
\begin{itemize}
\item $C_{\sliced E}$ contains disequalities $t_1\noteq t_2$ such
  that $t_1$ and $t_2$ have the same E-graph representatives;
  such disequalities are false because of the equalities in $\sliced E$.
\item $C_{\mathcal M}$ contains disequalities $t_1\noteq t_2$ such
  that $t_1$ and $t_2$ have distinct representatives $t'_1$ and $t'_2$
  with $\sem{\mathcal M}{t'_1}=\sem{\mathcal M}{t'_2}$;
  these are false because of $\mathcal M$.
\item $C_{\sf interface}$ contains disequalities $t_1 \noteq t_2$
  such that $t_1$ and $t_2$ have distinct representatives $t_1'$ and $t_2'$,
  $t_1'$ is evaluable and $t_2'$ is a slice;
  we can still satisfy $t_1\noteq t_2$ by picking a good value for $y$; we
  say $t_1'$ is an \emph{interface term}.
\item $C_{\sf free}$ contains disequalities $t_1\noteq t_2$ such that
  $t_1$ and $t_2$ have distinct slices as representatives;
  we can still satisfy $t_1\noteq t_2$ by picking a good value for $y$.
\end{itemize}
The disjuncts in $\sliced D$ take part in the conflict either when (i)
one of the clauses in $\sliced D$ is false because $C_{\sf interface}$
and $C_{\sf free}$ are both empty; or (ii) the finite domains are too
small to satisfy the disequalities in $C_{\sf interface}$ and $C_{\sf free}$,
given the values assigned in $\mathcal M$.
In either case, we  can produce a conflict explanation with
Algorithm~\ref{algo:disconflict}.

\begin{algorithm}[!t]
  \caption{Disequality conflict}\label{algo:disconflict}
  \begin{algorithmic}[1]
    \Function{dis\_conflict}{$\sliced D,\mathcal M,\mathcal G$}
    \State $S\gets\emptyset$
    \Comment{where we collect interface terms}
    \State $C_0\gets\emptyset$
    \Comment{where we collect the disequalities that evaluate to false}
    \For{$C\in \sliced D$}
    \State{$C_{\mathcal M}^{\sf rep}\gets \bigvee
      \{\Call{rep}{t_1,\mathcal G}
      \noteq
      \Call{rep}{t_2,\mathcal G}
      \mid
      (t_1\noteq t_2) \in C_{\mathcal M}
      \}$}
    \If{$\Call{is\_empty}{C_{\sf interface}}$ and $\Call{is\_empty}{C_{\sf free}}$}
    \State{${\sf raise\_conflict}(E\wedge D\implies C^{\sf rep}_{\mathcal M})$}
    \Else
    \State{$C_0\gets C_0\vee  C^{\sf rep}_{\mathcal M}$}
    \Comment{we collect the disequalities made false in the model}
    \For{$t_1\noteq t_2\in C_{\sf interface}$ with
      $\evaluates[\mathcal M]{\Call{rep}{t_1,\mathcal G}}$
    }
    \State{$S\gets S\cup\{\Call{rep}{t_1,\mathcal G}\}$}
    \Comment{we collect the interface term}
    \EndFor
    \EndIf
    \EndFor
    \State
    $C_{\neq}\gets \bigvee \{ t_1\eq t_2
    \mid \eval[\mathcal M]{t_1}\neq\eval[\mathcal M]{t_2},\
    t_1,t_2\in S \}$
    \State
    $C_=\gets \bigvee \{ t_1\noteq t_2
    \mid \eval[\mathcal M]{t_1}=\eval[\mathcal M]{t_2},\
    t_1\neq t_2,\ t_1,t_2\in S \}$
    \State\Return{$E\wedge D\implies C_0\vee C_{\neq} \vee C_=$}
    \EndFunction
  \end{algorithmic}
\end{algorithm}

In a type (i) conflict, the algorithm produces an interpolant $C^{\sf
  rep}_{\mathcal M}$ that is derived from a single element of
$\sliced D$.  Because we assume that $\mathcal{C}$ is a core, a type
(i) conflict can happen only if $\sliced D$ is a singleton.  Here is
how the algorithm behaves on such a conflict:
\begin{example}
  Let $r_1$ and $r_2$ be bit ranges of the same length,
  let $r_3$, $r_4$, $r_5$ be bit ranges of the same length.
  Assume $\sliced E$ contains
  \[\{\ \extract{x_1}{r_1}\eq\extract{y}{r_1}\ ,\
  \extract{x_2}{r_2}\eq\extract{y}{r_2}\ ,\
  \extract{y}{r_3}\eq\extract{y}{r_5}\ ,\
  \extract{y}{r_4}\eq\extract{y}{r_5}\ \},\]
  and assume $\sliced D$ is the singleton
  $\{\ (\extract{y}{r_1}\noteq\extract{y}{r_2}
  \vee
  \extract{y}{r_3}\noteq\extract{y}{r_4})\
  \}$.
  Let $\mathcal M$ map $x_1$ and $x_2$ to $0\ldots 0$
  and assume $\extract{y}{r_5}$ is the E-graph representative for component
  \[\{\ \extract{y}{r_3}, \extract{y}{r_4}, \extract{y}{r_5}\ \}.\]
  The unique clause of $\sliced D$ contains two disequalities:
  \begin{itemize}[noitemsep,topsep=0pt]
  \item The first one, $\extract{y}{r_1}\noteq\extract{y}{r_2}$, belongs to $C_{\mathcal M}$
    because the representatives of $\extract{y}{r_1}$ and $\extract{y}{r_2}$,
    namely $\extract{x_1}{r_1}$ and $\extract{x_2}{r_2}$, both evaluate to $0\ldots0$.
  \item
    The second one, $\extract{y}{r_3}\noteq\extract{y}{r_4}$ ,belongs to $C_{\sliced E}$ because
    the representatives of $\extract{y}{r_3}$ and $\extract{y}{r_4}$
    are both $\extract{y}{r_5}$,
  \end{itemize}
  As $C_{\sf interface}$ and $C_{\sf free}$ are empty,
  Algorithm~\ref{algo:disconflict} outputs
  $E\wedge D \implies \extract{x_1}{r_1}\noteq\extract{x_2}{r_2}$.
\end{example}

%% in a type (i) conflict, $\sliced D$ is always a singleton: a clause of
%% $\sliced D$, together with $\sliced E$ and $\mathcal M$, empties the
%% range of feasible values for $y$, so having assumed that $E,D$ is a
%% core, $\sliced D$ must only contain that clause.

For a conflict of type (ii), the equalities and disequalities that
hold in $\mathcal M$ between the interface terms
make the slices of
$y$ require more values than there  exist. So the produced conflict
clause includes (the negation of) all such equalities and
disequalities.  An example can be given as follows:

\begin{example}
  Assume $E$ (and then $\sliced E$) is empty and assume $\sliced D$ is
  \[\{\ \extract{x_2}{0}\noteq\extract{x_2}{1}
  \vee
  \extract{y}{0}\noteq\extract{y}{1}\ ,\
  \extract{x_1}{0}\noteq\extract{y}{0}\ ,\
  \extract{x_1}{1}\noteq\extract{y}{1}\
  \}\]
  Let $\mathcal M$ map $x_1$ and $x_2$ to $00$.
  Then \Call{dis\_conflict}{$\sliced D,\mathcal M,\mathcal G$} behaves as follows:
  \begin{itemize}[noitemsep,topsep=0pt]
  \item In the first clause, call it $C$,
    the first disequality is in $C_{\mathcal M}$,
    as the two sides are in different components but evaluate to the same value;
    so $C_0$ becomes $\{\ \extract{x_2}{0}\noteq\extract{x_2}{1}\ \}$;
    the second disequality features two slices and is thus in $C_{\sf free}$;
    The clause is potentially satisfiable and we move to the next clause.
  \item The second clause contains a single disequality that cannot be evaluated (since $\extract{y}{0}$ is not evaluable in $\mathcal{M}$).
    Term $\extract{x_1}{0}$ is added to $S$.
    The clause is potentially satisfiable so we move to the next clause.
  \item The third clause of $\sliced D$ is similar. It contains a single disequality that cannot be evaluated.
    The interface term $\extract{x_1}{1}$ is added to $S$.
  \end{itemize}
  Since all clauses of $\sliced D$ have been processed, the conflict
  is of type (ii).  Indeed, $\extract{y}{0}$ must be different from
  $0$ because of the second clause, $\extract{y}{1}$ must also be
  different from $0$ because of the third clause, but $\extract{y}{0}$
  and $\extract{y}{1}$ must be different from each other because of
  the first clause.  Since both $\extract{y}{0}$ and $\extract{y}{1}$
  have only one bit, there are only two possible values for these two
  slices, so the three constrains are in conflict.
  Algorithm~\ref{algo:disconflict} produces the conflict clause
  \[D \implies
  (\ \extract{x_2}{0}\noteq\extract{x_2}{1}
  \vee
  \extract{x_1}{0}\noteq\extract{x_1}{1}\ ).\]
  The disequality $\extract{x_2}{0}\noteq\extract{x_2}{1}$ is necessary because,
  if it were true in $\mathcal M$,
  we would not have to satisfy $\extract{y}{0}\noteq\extract{y}{1}$
  and therefore $y\gets 11$ would work.
  Disequality $\extract{x_1}{0}\noteq\extract{x_1}{1}$ is also necessary because,
  if it were true in $\mathcal M$, say with $x_1\gets 01$ (resp.\ $x_1\gets 10$),
  then $y\gets 11$ (resp.\ $y\gets 00$) would work.
\end{example}

%% This was a minimalistic example for a conflict of type (ii)
%% where the conflict clause contains exactly what we need.
%% In general though,
%% we need to make sure that the conflict clause is indeed valid
%% and hopefully is not too much bigger than necessary.
%% To show that it is valid,
%% we show that the existence of a counter-model leads to the following contradiction.

Correctness of the method relies on the following lemma\longv[, whose proof can be found in~\cite{GrahamLengrandJovanovicDutertre:arXiv2020}]{}.
\begin{toappendix}
  \begin{lemma}[The produced clauses are interpolants]\strut
    \label{lem:concatextract}
    \begin{enumerate}
    \item If Algorithm~\ref{algo:egraph} reaches line 7,
      $t'_1\eq t'_2$ is an interpolant for $E\wedge D$ at $\mathcal M$.
    \item If Algorithm~\ref{algo:disconflict} reaches line 7,
      $C^{\sf rep}_{\mathcal M}$ is an interpolant for $E\wedge D$ at $\mathcal M$.
    \item\label{point:three}
      If it reaches line 14,
      $C_0\vee C_{\neq} \vee C_=$ is an interpolant for $E\wedge D$ at $\mathcal M$.
    \end{enumerate}
  \end{lemma}
\end{toappendix}

\longv{
  \begin{toappendix}[\begin{proof}See Appendix \thisappendix.\end{proof}]
    \begin{proof}
      The first two parts are straightforward. We prove point~\ref{point:three}.
      \begin{itemize}
      \item Free variables.\\
        By construction, $C_0$ has free variables in $\vec x$ (l.~5,~9).
        So does $S$ (l.11), and therefore $C_{\neq}$ and $C_{=}$.
      \item Validity.\\
        We show that $(E\wedge D)\implies(C_0\vee C_{\neq} \vee C_=)$ is valid.
        Let $\mathcal M'$ be a model for $\vec x,y$
        satisfying $E\wedge D$ but not $C_0\vee C_{\neq} \vee C_=$.
        Since $\mathcal M'$ satisfies $E\wedge D$,
        it satisfies $\sliced E$ and $\sliced D$,
        so for each component of the E-graph $\mathcal G$,
        $\mathcal M'$ evaluates each term of the component to the same value.
        And moreover it satisfies each clause $C$ in $\sliced D$.
        Take such a clause $C$:
        $\mathcal M'$ still evaluates $C_{\sliced E}$ to false
        because $\mathcal M'$ evaluates each term of a $\mathcal G$-component
        to the same value.
        As $\mathcal M'$ does not satisfy $C_0\vee C_{\neq} \vee C_=$
        it surely does not satisfy $C_0$.
        By construction (l.~5,~9) the disequalities in $C_0$
        are between representatives of disequalities in $C_{\mathcal M}$,
        so $\mathcal M'$ surely does not satisfy $C_{\mathcal M}$ either.
        So $\mathcal M'$ must satisfy $C$
        by satisfying a disequality $d_C$ in $C_{\sf interface}$ or $C_{\sf free}$.
        Moreover $\mathcal M'$ does not satisfy $C_{\neq} \vee C_=$
        and therefore for two interface terms $t_1$ and $t_2$,
        $\eval[\mathcal M']{t_1}=\eval[\mathcal M']{t_2}$ \iff\
        $\eval[\mathcal M]{t_1}=\eval[\mathcal M]{t_2}$,
        by construction of $C_{\neq}$ (l.~12) and $C_=$ (l.~13).
        Let $t_1,\ldots,t_m$ be the interface terms,
        with values $v_1,\ldots,v_m$ in $\mathcal M$
        and values $v'_1,\ldots,v'_m$ in $\mathcal M'$.
        Let $\pi$ be a sort-preserving permutation on all bitvector values that maps $v'_i$ to $v_i$ for $1\leq i\leq m$.
        Let us extend $\mathcal M$ by assigning to $y$ a value $v$ such that
        for each slice $\extract y r$,
        we have
        $\eval[\mathcal M,y\mapsto v]{\extract y r}=\pi(\eval[\mathcal M']{\extract y r})$.
        We know that $\mathcal M,y\mapsto v$ satisfies $\sliced E$, and therefore $E$.
        We now show that $\mathcal M,y\mapsto v$ satisfies $\sliced D$, and therefore $D$,
        by showing that for each clause $C$ in $\sliced D$ it satisfies $d_C$.
        Let $\extract y r$ and $t$ be the representatives of the two sides of $d_C$.
        Whether $t$ is a slice of $y$ ($d_C\in C_{\sf free}$)
        or is an interface term ($d_C\in C_{\sf interface}$)
        we have in both cases $\pi(\eval[\mathcal M']{t})=\eval[\mathcal M,y\mapsto v]{t}$.
        Since $\mathcal M'$ satisfies $\sliced E$ and $d_C$,
        we have $\eval[\mathcal M']{\extract y r}\neq\eval[\mathcal M']{t}$,
        and therefore
        $\eval[\mathcal M,y\mapsto v]{\extract y r}\neq\eval[\mathcal M,y\mapsto v]{t}$.
        Since $\mathcal M,y\mapsto v$ satisfies $\sliced E$, it satisfies $d_C$.
      \item Falsification by $\mathcal M$.\\
        By construction, $\mathcal M$ falsifies $C_{\neq}$ (l.~12) and $C_=$ (l.~13).
        Moreover each disequality in $C_0$
        is between the representatives of a disequality $C_{\mathcal M}$
        for some clause $C\in\sliced D$ (l.~5,~9).
        Since $\mathcal M$ falsifies $C_{\mathcal M}$ by definition,
        and satisfies $\sliced E$ (otherwise Algorithm~\ref{algo:egraph}
        would have raised a conflict),
        $\mathcal M$ also falsifies the disequality between the representatives.
        So $\mathcal M$ falsifies $C_0$.
      \end{itemize}

      %% The second question is whether a smaller clause could be used as a valid clause.
      %% Indeed, placing in the conflict clause
      %% the negation of all equalities and disequalities that hold between the interface terms
      %% is somewhat of an overkill,
      %% if only because symmetry and transitivity of equality makes that clause a bit redundant.
      %% But also because we are actually trying to explain, with a clause,
      %% how to repair an unfeasible graph coloring problem:
      %% when thinking of the E-graph components as vertices
      %% and stated disequalities as edges between vertices,
      %% we are trying to extend the coloring imposed by $\mathcal M$
      %% into a coloring of the whole graph.
      %% As we run out of colors, it is likely that
      %% a more compact clause could describe what is wrong with $\mathcal M$'s coloring.
      %% However, the cost of computing a more minimal conflict clause might not be worth
      %% the efficiency gain that the smaller conflict clause would give.
      %% [SGL: to be tested?]
    \end{proof}
  \end{toappendix}
}

% !TEX root = main.tex
% !TEX program = pdflatex

\section{A Linear Arithmetic Fragment}
\label{sec:bvarith}

Our second specialized explanation mechanism applies when constraints
$\mathcal{C} = \lbrace C_1,\ldots, C_m \rbrace$ belong to the following grammar:
\[
\begin{array}{l@{\quad}llllll}
  \mbox{Constraints} & C &\recdef &a \sep \neg a &\\
  \mbox{Atoms} & a &\recdef
  & e_1 + t \uleq e_2 + t
  \sep e_1 \uleq e_2 + t
  \sep e_1 + t \uleq e_2\\
  \mbox{Terms} & t &\recdef
  & \bextract h y
  \sep \uextract l t
  \sep t + e_1 \sep - t
  \sep 0_k \circ t
  \sep t \circ 0_k
\end{array}
\]
where $e_1$ and $e_2$ range over \emph{evaluable} bitvector terms (\ie
$y\not\in\fv {e_1}\cup\fv {e_2}$), and $0_k$ is $0$ on $k$ bits. We
can represent variable $y$ as the term $\bextract {\bwidth y} y$. This
fragment of bitvector arithmetic is \emph{linear} in $y$ and there can
be only one occurrence of $y$ in terms. Constraints in
Section~\ref{sec:concatextract} are then outside this fragment in
general.

Let $\mathcal A$ be $\exists y (C_1\wedge \cdots \wedge C_m )$, and ${\mathcal
M}$ be the partial model involved in the conflict. The interpolant for $\mathcal
A$ at model ${\mathcal M}$ is (roughly) produced as follows:
\begin{enumerate}
\item
  For each constraint $C_i$, $1\leq i\leq m$,
  featuring a (necessarily unique) lower-bits extract $\bextract{w_i}y$,
  we compute a \emph{condition cube} $c_i$ satisfied by $\mathcal M$
  and a \emph{forbidden interval} $I_i$ of the form $\interval{l_i}{u_i}$,
  where $l_i$ and $u_i$ are evaluable terms,
  such that $c_i \implies (C_i\equivalent (\bextract{w_i}y\notin I_i))$ is valid.
\item
  We group the resulting intervals $(I_i)_{1\leq i\leq m}$ according to their bitwidths:
  if $\mathcal S_w$ is the set of intervals forbidding values for $\bextract{w}y$,
  $1\leq w\leq \bwidth y$,
  then under condition $\bigwedge_{i=1}^m c_i$
  formula $\mathcal A$ is equivalent to
  $\exists y (\bigwedge_{w=1}^{\bwidth y}\ (\ \bextract{w}y\notin \bigcup_{I\in\mathcal S_w} I\ ))$.
\item We produce a series of constraints $d_1$,\ldots, $d_p$
  that are satisfied by $\mathcal M$ and that are inconsistent with
  $\bigwedge_{w=1}^{\bwidth y}\ (\ \bextract{w}y\notin \bigcup_{I\in\mathcal S_w} I\ )$.
  The interpolant will be $(\bigwedge_{i=1}^m c_i \wedge \bigwedge_{i=1}^p d_i) \imp \bot $:
  it is implied by $\mathcal A$, and evaluates to false in ${\mathcal M}$.
\end{enumerate}

\subsection{Forbidden Intervals}\label{sec:intervals}

An \emph{interval} takes the form $\interval l u$,
where the lower bound $l$ and upper bound $u$ are evaluable terms of some bitwidth $w$,
with $l$ included and $u$ excluded.
The notion of interval used here is considered modulo $2^w$. We do not require $l \leq^uu$ so an interval may ``wrap around'' in $\ZZ/2^w\ZZ$.
For instance, the interval $\interval {1111}{0001}$ contains two bitvector values, namely, $1111$ and $0000$.
If $l$ and $u$ evaluate to the same value, then we consider $\interval l u$ to be empty (as opposed to the full domain, which we
denote by $\full[w]$ or just $\full$).
Notation $t\in I$ stands for literal $\top$ if $I$ is $\full$ and literal $t{-}l <^u u{-}l$ if $I$ is $\interval l u$.
The value in model $\mathcal M$ of an evaluable term $e$ (\resp evaluable cube $c$, interval $I$)
is denoted $\sem{\mathcal M}{e}$ (\resp $\sem{\mathcal M}{c}$, $\sem{\mathcal M}{I}$).

\begin{table}[t]
\[\small
\begin{array}{|c|c|c|c|c|c|}
\hline
\multirow{2}{*}{\textbf{Atom } a}
& \multicolumn3{c|}{\textbf{Forbidden interval that $a$ (\resp $\neg a$) specifies for $t$}}& \multirow{2}{*}{}\\\cline{2-4}
  & I_{a} & I_{\neg a} &\mbox{\textbf{Condition } $c_a$/$c_{\neg a}$}&\\\hline\hline

\multirow{2}{*}{$e_1+t\uleq e_2+t$} & \interval{-e_2}{-e_1} & \interval{-e_1}{-e_2} & e_1\noteq e_2 &{\textcolor{gray}1} \\\cline{2-5}

& \empt & \full & e_1\eq e_2 & {\textcolor{gray}2}\\\hline

\multirow{2}{*}{$e_1\uleq e_2+t$} & \interval{-e_2}{e_1-e_2} & \interval{e_1-e_2}{-e_2} & e_1\noteq 0 & {\textcolor{gray}3}\\\cline{2-5}

& \empt & \full & e_1\eq 0 &{\textcolor{gray}4}\\\hline

\multirow{2}{*}{$e_1+t\uleq e_2$} & \interval{e_2-e_1+1}{-e_1} & \interval{-e_1}{e_2-e_1+1} & e_2 \noteq -1 & {\textcolor{gray}5}\\\cline{2-5}

& \empt& \full & e_2 \eq -1 &{\textcolor{gray}6}\\\hline

%% \multirow{2}{*}{$e_1\uleq e_2$} & \full & \empt & e_2 \ult e_1 &{\textcolor{gray}7}\\\cline{2-5}

%% & \empt & \full & e_1 \uleq e_2 &{\textcolor{gray}8}\\\hline

\end{array}
\]
\caption{Creating the forbidden intervals}
\label{tab:cases}
\end{table}

Given a constraint $C$ with unevaluable term $t$,
we produce an interval $I_C$ of forbidden values for $t$
according to the rules of Table~\ref{tab:cases}.
A side condition literal $c_C$ identifies when the lower and upper bounds
would coincide, in which case the interval produced is either empty or
full. For every row of the table, the formula $c_C \implies ( C
\equivalent t\notin I_C)$ is valid in \BVth.  Given a partial model
$\mathcal{M}$, we convert $C$ to such an interval by selecting the row
where $\sem{\mathcal M}{c_C}=\vtrue$.

%% By applying these rules to a constraint $C$, we construct an
%% interval $I_C$ and a side-condition $c_C$ such that 
%% We say that \emph{$p$ disambiguates $I$} if $I$ is $\full$, $I$ is $\empt$,
%% or $I=\interval l u$ with $p\implies l\noteq u$ valid.
%% The last two cases concern the situation where $y$ does not appear in the constraint.
%% As explained above for constraints with a different bitwidth than $w$,
%% constraints without $y$ could also be taken out of formula $\mathcal A$.
%% The interpolation mechanism looks at which line of the table applies,
%% so that $\sem{\mathcal M}{c_C}=\vtrue$,
%% and outputs the corresponding interval $I_C$.
%% In each case, $c_C$ disambiguates $I_C$
%% and $c_C \implies ( C \equivalent t\notin I_C)$ is valid.

\begin{example}\label{ex:1}\strut
  \begin{enumerate}
  \item[\ref{ex:1}.1] % Taken from tests/regress/mcsat/bv/bench_962.smt2.dd.smt2
    Assume $C_1$ is literal $\neg(x_1\uleq y)$ and ${\mathcal M}=\{x_1\mapsto 0000\}$.
    Then line 4 of Table~\ref{tab:cases} applies,
    and $I_{C_1}$ is interval $\full$ with condition $x_1\eq 0$.
  \item[\ref{ex:1}.2]
    Assume  $C_1$ is $\neg(y \eq x_1)$, $C_2$ is $(x_1 \uleq x_3 + y)$, $C_3$ is $\neg(y - x_2 \uleq x_3 + y)$,
    and ${\mathcal M}=\{x_1\mapsto 1100,x_2\mapsto 1101,x_3\mapsto 0000\}$.
    Then by line 5, $I_{C_1} = \interval{x_1}{x_1+1}$ with trivial condition $(0\noteq -1)$,
    by line 3, $I_{C_2} = \interval{-x_3}{x_1-x_3}$ with condition $(x_1\noteq 0)$,
    and by line 1, $I_{C_3} = \interval{x_2}{-x_3}$ with condition $(-x_2\noteq x_3)$.
  \end{enumerate}
\end{example}

\begin{figure}[t]
  \small
  \begin{tabular}{|c|}
    \hline
    \parbox[t]{\textwidth}{
      \[[-5]
      \begin{array}{lll@{\ \eqdef\ }l@{\qquad}l@{\ \eqdef\ }l}
        \Iinverse{t}{&\empt} {&c} & (1,\ \empt,\ c)
        & \Iinverse{0_k \circ t}{I} c & \Iuptrim k {t}{I} c\\
        \Iinverse{t}{&\full} {&c} & (1,\ \full,\ c)
        & \Iinverse{t \circ 0_k}{I} c & \Idowntrim k {t}{I} c\\
        \Iinverse{\bextract w y}{&I} {&c} & (w,\ I,\ c)
        & \multicolumn2{r}{\mbox{when $I$ is not $\empt$ nor $\full$}}\\
        \Iinverse{\uextract w t}{&\interval l u} {&c} & \multicolumn3{l}{\Iinverse t{\interval {l\circ 0_w} {u\circ 0_w}} c}\\
        \Iinverse{t + c}{&\interval l u}{&c}          & \multicolumn3{l}{\Iinverse t{\interval {l{-}c} {u{-}c}} c}\\
        \Iinverse{- t}{&\interval l u}{&c}           & \multicolumn3{l}{\Iinverse t{\interval {1{-}u} {1{-}l}} c}
      \end{array}
      \]
    }
    \\\hline
    \parbox[t]{\textwidth}{
      \[[-5]
      \begin{array}{l@{}l}
        \multirow{3}{*}{$\Iuptrim k {t}{\interval l u} c \eqdef\!\left\{\rule[-12pt]{0cm}{12pt}\right.$}
         & \Iinverse t{\interval {l'} {u'}} {c{\wedge} c_l{\wedge} c_u}
        \hfill\mbox{if $\interval {l'} {u'}$ is not $\empt$} \\
         & (1,\ \full,\ c{\wedge} c_l{\wedge} c_u {\wedge} c')
        \hfill\mbox{if $\interval {l'} {u'}$ is $\empt$ and $\sem{\mathcal M}{c'}$ is true} \\
         & (1,\ \empt,\ c{\wedge} c_l{\wedge} c_u {\wedge} \neg c')
        \ \ \mbox{if $\interval {l'} {u'}$ is $\empt$ and $\sem{\mathcal M}{c'}$ is false}\\
      \end{array}
      \]
      \begin{tabular}{lllll@{\qquad}l}
        where
        & $l'$ is $\bextract w l$ & (\resp $0_w$) & and $c_l$ is $a_l$ & (\resp $\neg a_l$) & if $\sem{\mathcal M} {a_l}$ is true\hfill (\resp false),\\
        & $u'$ is $\bextract w u$ & (\resp $0_w$) & and $c_u$ is $a_u$ & (\resp $\neg a_u$) & if $\sem{\mathcal M} {a_u}$ is true (\resp false),\\
        & \multicolumn5{l}{$a_l$ is $\uextract w l \eq 0_k$,\qquad
        $a_u$ is $\uextract w u \eq 0_k$,\qquad
        $c'$ is $(0_{k+w}\in\interval l u)$,\qquad and
        $w$ is $\bwidth t$.}
      \end{tabular}
      \medskip
    }
    \\\hline
    \parbox[t]{\textwidth}{
      \[[-5]
      \begin{array}{l@{}l}
        \multirow{3}{*}{$\Idowntrim k {t}{\interval l u} c \eqdef\!\left\{\rule[-12pt]{0cm}{12pt}\right.$}
         & \Iinverse t{\interval {l'} {u'}} {p{\wedge} c_l{\wedge} c_u}
        \hfill\mbox{if $\interval {l'} {u'}$ is not $\empt$} \\
        & (1,\ \full,\ c{\wedge} c_l{\wedge} c_u {\wedge} c')
        \hfill\mbox{if $\interval {l'} {u'}$ is $\empt$ and $\sem{\mathcal M}{c'}$ is true} \\
        & (1,\ \empt,\ c{\wedge} c_l{\wedge} c_u {\wedge} \neg c')
        \ \ \mbox{if $\interval {l'} {u'}$ is $\empt$ and $\sem{\mathcal M}{c'}$ is false}\\
      \end{array}
      \]
      \begin{tabular}{lllll@{\ }l}
        where
        & $l'$ is $\uextract k l$ & (\resp $\uextract k l{+}1$)
        & and $c_l$ is $a_l$ & (\resp $\neg a_l$)
        & if $\sem{\mathcal M} {a_l}$ is true\hfill  (\resp false),\\
        & $u'$ is $\uextract k u$ & (\resp $\uextract k u{+}1$)
        & and $c_u$ is $a_u$ & (\resp $\neg a_u$)
        & if $\sem{\mathcal M} {a_u}$ is true (\resp false),\\
        & \multicolumn5{l}{$a_l$ is $\bextract k l \eq 0_k$,\qquad
        $a_u$ is $\bextract k u \eq 0_k$,\qquad
        $c'$ is $(u' \circ 0_k\in\interval l u)$,\qquad and
        $w$ is $\bwidth t$.}
      \end{tabular}
      \medskip
    }
    \\\hline
  \end{tabular}
  \caption{Transforming the forbidden intervals}
  \label{fig:Iinverse}
\end{figure}
Given the supported grammar, term $t$ contains a unique subterm of the form
$\bextract{w}y$. We transform $I_C$ into an interval of forbidden
values for $\bextract{w}y$ by applying procedure
$\Iinverse{t}{I_C}{c_C}$ shown in Figure~\ref{fig:Iinverse},
which proceeds by recursion on $t$.
Its specification is given below, and correctness is proved by induction on $t$.
\begin{lemma}[Correctness of forbidden intervals]
  Assuming cube $c$ is true in $\mathcal{M}$,
  then $\Iinverse{t}{I} {c}$ returns a triple $(w, I', c')$ such that
  $c'$ is a cube that is true in $\mathcal{M}$,
  and both $c'\implies c$ and $c'\implies (t\notin
  I\equivalent\bextract {w} y\notin I')$ are valid in \BVth.
\end{lemma}
%% This is done by calling $\Iinverse{t}{I_C} {c_C}$, which proceeds by recursion on $t$, following the grammar, as shown in Fig.~\ref{fig:Iinverse}.
%% A generic call $\Iinverse{t}{I} {c}$ assumes that
%% $c$ disambiguates $I$ of bitwidth $w {=} \bwidth t$
%% and $\sem{\mathcal M}c$ is true;
%% it outputs a triple $(w', I', c')$ where
%% $c'$ disambiguates $I'$ of bitwidth $w'$,
%% $\sem{\mathcal M}{c'}$ is true,
%% and both $c'\implies c$ and $c'\implies (t\notin I\equivalent\bextract {w'} y\notin I')$ are valid.
%% Argument $c$ works as an accumulator of facts about $\mathcal M$.
Running $\Iinverse{t_{C_i}}{I_{C_i}} {c_{C_i}}$ for all constraints $C_i$, $1{\leq} i{\leq} m$,
produces a family of triples $(w_i,I'_i,c'_i)_{1\leq i\leq m}$ such that, for each $i$,
formula $c'_i\implies (C_i \equivalent (\bextract {w_i} y\notin I'_i))$ is valid in \BVth\ and $c'_i$ is true in $\mathcal M$.

\subsection{Interpolant}\label{sec:interp}

First, assume that one of the triples obtained above
is of the form $(w,\full,c)$, coming from constraint $C$.
As the interval forbids the full domain of values for $\bextract w y$,
we produce conflict clause $C\wedge c\implies\bot$.
This formula is an interpolant for $\mathcal A$ at $\mathcal M$.
This is illustrated in Example~\ref{ex:2}.1.
\begin{example}\label{ex:2}\strut
  \begin{enumerate}
  \item[\ref{ex:2}.1] In Example~\ref{ex:1}.1 where $C_1$ is literal $\neg(x_1\uleq y)$ and ${\mathcal M}=\{x_1\mapsto 0000\}$,
  the interpolant for $\neg(x_1\uleq y)$ at ${\mathcal M}$ is $(x_1\eq 0)\imp\bot$.
\item[\ref{ex:2}.2] Example~\ref{ex:1}.2 does not contain a full
  interval.  Model ${\mathcal M}$ satisfies the three conditions
  $c_1\eqdef(0\noteq -1)$, $c_2\eqdef(x_1\noteq 0)$ and
  $c_3\eqdef(-x_2\noteq x_3)$, and the intervals $I_{1} =
  \interval{x_1}{x_1+1}$, $I_{2} = \interval{-x_3}{x_1-x_3}$, and
  $I_{3} = \interval{x_2}{-x_3}$, evaluate to $\sem{\mathcal M}{I_1} =
  \interval{1100}{1101}$, $\sem{\mathcal M}{I_2} =
  \interval{0000}{1100}$, and $\sem{\mathcal M}{I_3} =
  \interval{1101}{0000}$, respectively.  Note how
  $\bigcup_{i=1}^3\sem{\mathcal M}{I_i}$ is the full domain.
  \end{enumerate}
\end{example}

\begin{figure}[t]
\[
\begin{array}{|c|c@{\ }c@{\ }c@{}c@{}c|}
  \hline
  \mbox{\bf bitwidth} & w_1 & > & w_2 & > \cdots > & w_j\\\hline
  \mbox{\bf Interval layer} & \mbox{$w_1$-intervals} && \mbox{$w_2$-intervals} & \ldots & \mbox{$w_j$-intervals}\\
  & \mathcal S_1=\{I_{1.1},I_{1.2},\ldots \} && \mathcal S_2=\{I_{2.1},I_{2.2},\ldots \} & \ldots & \mathcal S_j=\{I_{j.1},I_{j.2},\ldots \}\\\hline
  \mbox{\bf
    \begin{tabular}c
      Forbidding\\
      values for
  \end{tabular}} & \bextract{w_1}y && \bextract{w_2}y & \ldots & \bextract{w_j}y\\
  \hline
\end{array}
\]
\caption{Intervals collected from $C_1\wedge\cdots \wedge C_m$}
\label{fig:intervals}
\end{figure}
Assume now that no interval is full (as in Example~\ref{ex:2}.2).
%% We first remove from the family of triples those of the form $(w,\empt,p)$:
%% they do not affect the values that are feasible for $y$ in any way and,
%% together with the constraints from which they came,
%% can therefore be ignored in the production of the conflict clause.
We group the triples $(w,I,c)$ into different \emph{layers}
characterized by their bitwidths $w$:
$I$ will henceforth be called a \emph{$w$-interval},
restricting the feasible values for $\bextract w y$,
and $c_I$ denotes its associated condition in the triple.
Ordering the groups of intervals by decreasing bitwidths $w_1 > w_2 > \cdots > w_j$, as shown in Figure~\ref{fig:intervals},
$\mathcal S_j$ denotes the set of produced $w_j$-intervals.
The properties satisfied by the triples entail that
\[
\mathcal A \wedge (\bigwedge_{i=1}^j\bigwedge_{I\in\mathcal S_i}c_I) \implies \mathcal B
\]
is valid, where $\mathcal B$ is $\exists y \bigwedge_{i=1}^j (\bextract{w_i} y\notin \bigcup_{I\in\mathcal S_i} I)$.
And formula $(\bigwedge_{i=1}^j\bigwedge_{I\in\mathcal S_i}c_I) \implies \mathcal B$
is false in $\mathcal M$.
To produce an interpolant,
we replace $\mathcal B$ by a quantifier-free clause.
%% , stating the conjunction of all conditions implies
%% the equivalence between $\mathcal A$ and the conjunction of the restrictions on each $\bextract{w_i} y$:
%% \[
%% (\bigwedge_{i=1}^j\bigwedge_{I\in\mathcal S_i}c_I) \implies (\mathcal A \equivalent \bigwedge_{i=1}^j (\bextract{w_i} y\notin \bigcup_{I\in\mathcal S_i} I))
%% \]

The simplest case is when there is only one bitwidth $w=w_1$:
the fact that $\mathcal B$ is falsified by $\mathcal M$
means that $\bigcup_{I\in\mathcal S_1}\sem{\mathcal M}{I}$ is the full domain $\ZZ/2^w\ZZ$.
Property ``$\bigcup_{I\in\mathcal S_1}{I}$ is the full domain''
is then expressed symbolically
as a conjunction of constraints in the bitvector language.
%% expressing some property about model ${\mathcal M}$ that leaves no feasible value for $y$.
To compute them, we first extract
a sequence $I_1, \ldots, I_q$ of intervals from the set $\mathcal S_1$,
originating from a subset $\mathcal C$ of the original constraints $(C_i)_{i=1}^m$,
and such that the sequence $\sem{\mathcal M}{I_{1}},\ldots,\sem{\mathcal M}{I_q}$
of \emph{concrete} intervals leaves no ``hole'' between an interval of the sequence and the next,
and goes round the full circle of domain $\ZZ/2^w\ZZ$:
the sequence forms a circular chain of linking intervals.
This chain can be produced by a standard coverage extraction
algorithm\longv[ (see, e.g.,~\cite{GrahamLengrandJovanovicDutertre:arXiv2020})]{,
as shown in Appendix~\ref{app:simple}, Fig.~\ref{algo:sequence}}.
Formula $\mathcal B \eqdef \exists y (\bextract{w} y\notin \bigcup_{I\in\mathcal S_1} I)$ is then replaced by $(\bigwedge_{i=1}^q u_i\in I_{i+1})\implies \bot$,
where $u_i$ is the upper bound of $I_i$ and $I_{q+1}$ is $I_1$.
Each interval has its upper bound in the next interval ($u_i\in I_{i+1}$),
\ie intervals do link up with each other.
The conflict clause is then
\[
(\mathcal C
\wedge (\bigwedge_{i=1}^q c_{I_i})
\wedge (\bigwedge_{i=1}^q u_i\in I_{i+1}))
\implies \bot
\]
\begin{example}\label{ex:4}
  For Example~\ref{ex:2}.2,
  the coverage-extraction algorithm
  produces the sequence $I_1,I_3,I_2$,
  \ie $\interval{x_1}{x_1{+}1}, \ \interval{x_2}{-x_3}, \ \interval{-x_3}{x_1{-}x_3}$.
  The linking constraints are then
  $d_3\eqdef (x_1{+}1)\in I_3$, $d_2\eqdef({-}x_3)\in I_2$, and  $d_1\eqdef(x_1{-}x_3)\in I_1$,
  %% as the three constraints
  %% $d_3 = (x_1{+}1{-}x_2\ult-x_3{-}x_2)$, $d_2 = (0\ult x_1)$, and $d_1 = (-x_3\ult 1)$.
  and the interpolant is
  $d_3\wedge d_2\wedge d_1\imp\bot$.\footnote{We omit $c_1$, $c_2$, $c_3$ here,
    since they are subsumed by $d_1$, $d_2$, $d_3$, respectively.}
\end{example}

When several bitwidths are involved,
the intervals must ``complement each other'' at different bitwidths
so that no value for $y$ is feasible.
For a bitwidth $w_i$,
the union of the $w_i$-intervals in model ${\mathcal M}$
may not necessarily cover the full domain
(\ie $\bigcup_{I\in\mathcal S_i}\sem{\mathcal M}{I}$ may be different from $\ZZ/2^{w_{i}}\ZZ$).
The coverage can leave ``holes'',
and values in that hole are ruled out by constraints of other bitwidths.
To produce the interpolant, we adapt the coverage-extraction algorithm
into Algorithm~\ref{algo:sequence2},
which takes as input the sequence of sets $(\mathcal S_1,\ldots,\mathcal S_j)$ as described in Figure~\ref{fig:intervals},
and produces the interpolant's constraints $d_1,\ldots, d_p$,
collected in set $\textsf{output}$.
The algorithm proceeds in decreasing bitwidth order, starting with $w_1$,
and calling itself recursively on smaller bitwidths
to cover the holes that the current layer leaves uncovered
(termination of that recursion is thus trivial).
For every hole that $\bigcup_{I\in\mathcal S_1}\sem{\mathcal M}{I}$ leaves uncovered,
it must determine how intervals of smaller bitwidths can cover it.

\begin{algorithm}[t]
  \caption{Producing the interpolant with multiple bitwidths}\label{algo:sequence2}
  \scriptsize
  \begin{algorithmic}[1]
    \Function{cover}{$(\mathcal S_1,\ldots,\mathcal S_j),{\mathcal M}$}
    \State $\textsf{output}   \gets \emptyset$ \Comment{output initialized with the empty set of constraints}
    \State $\textsf{longest}  \gets \Call{longest}{\mathcal S_1,{\mathcal M}}$ \Comment{longest interval identified}
    \State $\textsf{baseline} \gets \textsf{longest.upper}$ \Comment{where to extend the coverage from}
    \While{$\sem{\mathcal M}{\textsf{baseline}} \not\in \sem{\mathcal M}{\textsf{longest}} $}
    \If{$\exists I\in \mathcal S_1, \sem{\mathcal M}{\textsf{baseline}} \in \sem{\mathcal M} I$}

    \State $I \gets \Call{furthest\_extend}{\textsf{baseline},\mathcal S_1,{\mathcal M}}$
    \State $\textsf{output}   \gets
    \textsf{output}
    \cup \{c_I, \textsf{baseline} \in I\}$
    \Comment{adding $I$'s condition and linking constraint}
    \State $\textsf{baseline} \gets I\textsf{.upper}$ \Comment{updating the baseline for the next interval pick}

    \Else\Comment{there is a hole in the coverage of $\ZZ/2^{w_1}\ZZ$ by intervals in $\mathcal S_1$}

    \State $\textsf{next} \gets \Call{next\_covered\_point}{\textsf{baseline},\mathcal S_1,{\mathcal M}}$ \Comment{the hole is $\interval{\textsf{baseline}}{\textsf{next}}$}
    \If{$\sem{\mathcal M}{\textsf{next}} - \sem{\mathcal M}{\textsf{baseline}} \ult 2^{w_2}$}
    \State $I \gets \interval{\bextract{w_2}{\textsf{next}}}{\bextract{w_2}{\textsf{baseline}}}$ \Comment{it is projected on $w_2$ bits and complemented}
    \State
    $\textsf{output} \gets \textsf{output}
    \cup \{\textsf{next} {-} \textsf{baseline} \ult 2^{w_2}\}
    \cup\Call{cover}{((\mathcal S_2\cup I),\mathcal S_3,\ldots,\mathcal S_j),{\mathcal M}}$
    \State $\textsf{baseline} \gets \textsf{next}$ \Comment{updating the baseline for the next interval pick}
    \Else\Comment{intervals of bitwidths $\leq w_2$ must forbid all values for $\bextract{w_2}y$}
    \State\Return{$\Call{cover}{(\mathcal S_2,\ldots,\mathcal S_j),{\mathcal M}}$} \Comment{$\mathcal S_1$ was not needed}
    \EndIf
    \EndIf
    \EndWhile
    \State\Return{$\textsf{output}\cup \{\textsf{baseline} \in \textsf{longest}\}$} \Comment{adding final linking constraint}
    \EndFunction
  \end{algorithmic}
\end{algorithm}

Algorithm~\ref{algo:sequence2} relies on the following ingredients:
\begin{itemize}
\item $\Call{longest}{\mathcal S,{\mathcal M}}$ returns an interval among $\mathcal S$ whose concrete version $\sem{\mathcal M}{I}$ has maximal length;
\item $I\textsf{.upper}$ denotes the upper bound of an interval $I$;
\item $\Call{furthest\_extend}{a,\mathcal S,{\mathcal M}}$
  returns an interval $I\in\mathcal S$
  that furthest\linebreak extends $a$ according to ${\mathcal M}$ (technically,
  an interval $I$ that $\uleq$-maximizes $\sem{\mathcal M}{I\textsf{.upper}-a}$
  among those intervals $I$ such that $\sem{\mathcal M}{a}\in \sem{\mathcal M}{I}$).
\item If no interval in $\mathcal S$ covers $a$ in $\mathcal M$,
  $\Call{next\_covered\_point}{a,\mathcal S, {\mathcal M}}$
  outputs the lower bound $l$ of an interval in $\mathcal S$
  that $\uleq$-minimizes $\sem{\mathcal M}{l-a}$.
\end{itemize}
Algorithm~\ref{algo:sequence2} proceeds by successively moving a concrete bitvector value
$\textsf{baseline}$ around the circle $\ZZ/2^{w_1}\ZZ$.
The baseline is moved when a symbolic reason why it is a forbidden value is found,
in a \textsf{while} loop that ends when the baseline has gone round the full circle.
If there is at least one interval in $\mathcal S_1$
that covers $\textsf{baseline}$ in ${\mathcal M}$ (l.\ 6),
the call to $\Call{furthest\_extend}{\textsf{baseline},\mathcal S_1,{\mathcal M}}$
succeeds, and $\textsf{output}$ is extended with condition $c_I$ and $(\textsf{baseline} \in I)$ (l.\ 8).
If not, a hole has been discovered, whose extent is given by
$\Call{next\_covered\_point}{\textsf{baseline},\mathcal S_1,{\mathcal M}}$ (l.\ 11).
If the hole is bigger than $2^{w_2}$
(\ie $2^{w_2}\uleq \sem{\mathcal M}{\textsf{next}{-}\textsf{baseline}}$),
then the intervals of layers $w_2$ and smaller
must rule out every possible value for $\bextract{w_2}y$,
and the $w_1$-intervals were not needed (l.\ 17).
If on the contrary the hole is smaller (\ie $\sem{\mathcal M}{\textsf{next}{-}\textsf{baseline}}\ult 2^{w_2}$),
then the $w_1$-interval $\interval {\textsf{baseline}}{\textsf{next}}$ is projected
as a $w_2$-interval
$I\eqdef\interval{{\bextract{w_2} {\textsf{baseline}}}}{{\bextract{w_2} {\textsf{next}}}}$
that needs to be covered by the intervals of bitwidth $w_2$ and smaller.
This is performed by a recursive call on bitwidth $w_2$ (l.\ 14);
the fact that only hole $I$ needs to be covered by the recursive call,
rather than the full domain $\ZZ/2^{w_{2}}\ZZ$,
is implemented by adding to $\mathcal S_2$ in the recursive call the complement
$\interval{{\bextract{w_2} {\textsf{next}}}}{{\bextract{w_2} {\textsf{baseline}}}}$
of $I$.
The result of the recursive call is added to the $\textsf{output}$ variable,
as well as the fact that the hole must be small.
The final interpolant is $(\bigwedge_{d\in\textsf{output}}d)\imp \bot$.
An example of run on a variant of Example~\ref{ex:1}.2 is given in\longv[~\cite{GrahamLengrandJovanovicDutertre:arXiv2020}]{Appendix~\ref{ex5}}.

%% \footnote{
%% Note that Algorithm~\ref{algo:sequence2} here
%% does not apply the same optimisation as Algorithm~\ref{algo:sequence} (line 9-11),
%% which detects when the use of interval $\textsf{longest}$ is unnecessary,
%% but it could also be done (at the cost of a more complex figure).}

\section{Normalization}\label{sec:normalisation}

\begin{figure}[t]
\[\small
\begin{array}{|l@{\ \leadsto\ }l@{\qquad}l@{\ }|r@{\ \leadsto\ }ll|}
  \hline

  u_1 \slt u_2   & \neg(u_2 \sleq u_1) &
  &
  u_1 \sleq u_2  & \multicolumn2{l|}{u_1{+}2^{\bwidth{u_1}{-}1} \uleq u_2{+}2^{\bwidth{u_2}{-}1}}\\

  u_1 \ult u_2   & \neg(u_2 \uleq u_1) &
  &
  u_1 \eq u_2  & u_1 - u_2 \uleq 0 &\\\hline

  \extract u h [l]&  \uextract l{\bextract h u}
  &&
  \bextract h {\uextract l u}& \uextract l{\bextract {h{+}l} u} &\\

  \uextract l {(u_1{\circ} u_2)}& \uextract {l{-}\bwidth{u_2}} {u_1} &\mbox{if $\bwidth{u_2}\leq l$}
  &
  \bextract h {(u_1{\circ} u_2)}& \bextract {h} {u_2} &\mbox{if $h\leq\bwidth{u_2}$}\\

  \uextract l {(u_1{\circ} u_2)}& {u_1} \circ \uextract {l} {u_2} &\mbox{if not}
  &
  \bextract h {(u_1{\circ} u_2)}& \bextract {h{-}\bwidth{u_2}} {u_1} \circ u_2&\mbox{if not}\\

  2^n\times u& \bextract {\bwidth u{-}n} u \circ 0_n&(n<\bwidth{u})
  &
  \bextract h {(u_1 {+} u_2)}& \bextract {h} {u_1} + \bextract {h} {u_2} & \\

  \bvnot u& -(u+1)&
  &
  \bextract h {(u_1 \times u_2)}& \bextract {h} {u_1} \times \bextract {h} {u_2} & \\

  \sextension k u& \multicolumn2{l|}{
    (0_k{\circ}(u{+}2^{\bwidth{u}{-}1})) {-} (0_k{\circ}2^{\bwidth{u}{-}1})
  }
  &
  \bextract h {({-}u)}& {-}\bextract {h} {u} &\\

  u_1 {\circ} u_2 & \multicolumn2{l|}{(u_1 {\circ} 0_{\bwidth {u_2}})+(0_{\bwidth {u_1}} {\circ} u_2)}
  &
  \multicolumn3{l|}{} \\
  
  \hline

\end{array}
\]
\caption{Rewriting rules}
\label{fig:rewriting}
\end{figure}
As implemented in Yices~2, \MCSAT processes a conflict by
first computing the conflict core with BDDs,
and then normalizing the constraints using the rules of Figure~\ref{fig:rewriting}.
In the figure, $u$, $u_1$ and $u_2$ stand for any bitvector terms,
$\sextension k u$ is the \emph{sign-extension} of $u$ with $k$ bits,
and $\bvnot u$ is the \emph{bitwise negation} of $u$.
The bottom left rule is applied with lower priority than the others
(as upper-bits extraction distributes over $\circ$ but not over $+$)
and only if exactly one of $\{u_1,u_2\}$ is evaluable (and not $0$).
In the implementation,
$\extract u {\bwidth u} [0]$ is identified with $u$,
$\circ$ is associative, and $+,\times$ are subject to ring normalization.
This is helped by the internal (flattened) representation of concatenations and bitvector polynomials in Yices~2.
Normalization allows the specialized interpolation procedure to apply at least to the following grammar:\footnote{$e_1 \lessdot e_2$ is accepted since it either constitutes the interpolant or it can be ignored.}
\[
\begin{array}{l@{\quad}llllll}
  \mbox{Atoms} & a &\recdef
  & e_1 + t \lessdot e_2 + t
  \sep e_1 \lessdot e_2 + t
  \sep e_1 + t \lessdot e_2
  \sep e_1 \lessdot e_2\\
  \mbox{Terms} & t &\recdef
  & \extract t h [l]
  \sep t + e_1 \sep - t
  \sep e_1 \circ t
  \sep t \circ e_1
  \sep \sextension k t
\end{array}
\]
where $\lessdot\in\{\uleq,\ult,\sleq,\slt,\eq\}$.
Rewriting can often help further,
by eliminating occurrences of the conflict variable (thus making more subterms evaluable)
and increasing the chances that two unevaluable terms $t_1$ and $t_2$ become syntactically equal in an atom
$e_1 {+} t_1 \lessdot e_2 {+} t_2$.\footnote{For this reason we normalize evaluable subterms of, e.g., $t_1$ and $t_2$.}
Finally, we cache evaluable terms to avoid recomputing conditions of the form $y\notin\fv e$.
These conditions are needed to determine whether the specialized procedures apply to a given conflict core.

% !TEX root = main.tex
% !TEX program = pdflatex

\section{Experiments}
\label{sec::experiments}

\begin{figure}[t]
  \centering
  \includegraphics[width=0.9\textwidth]{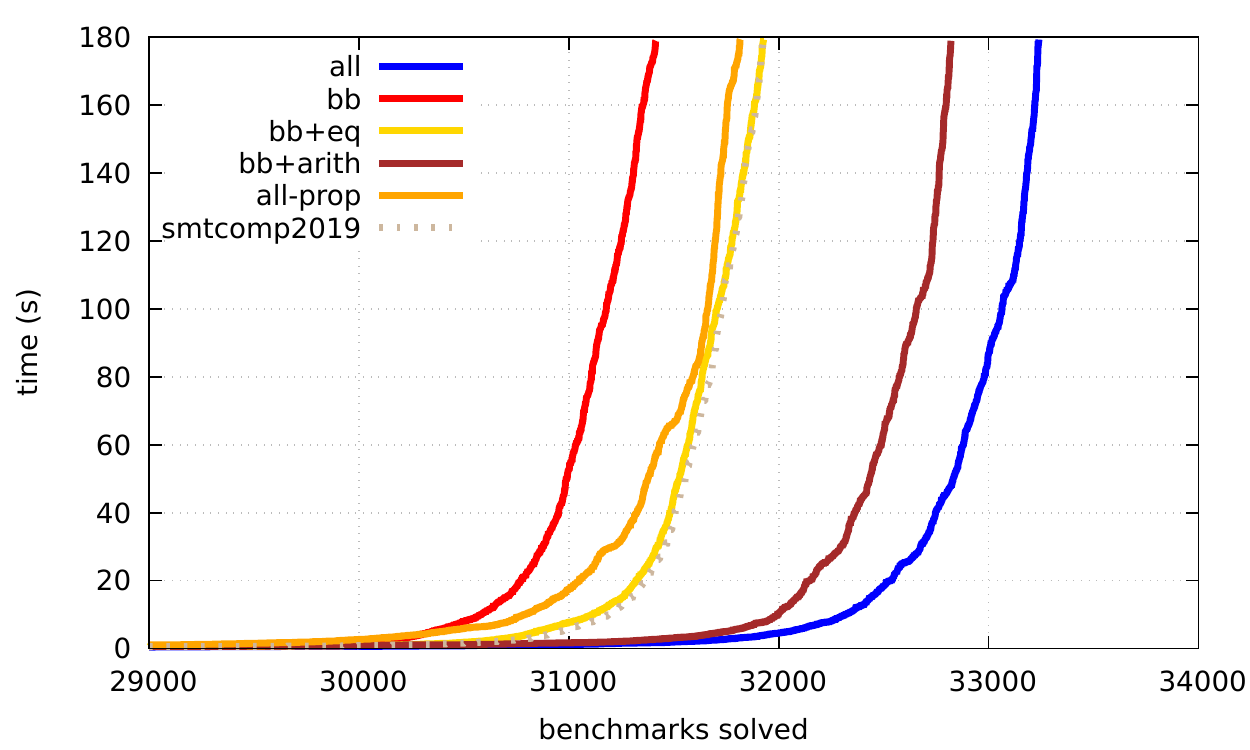}
  \caption{Evaluation of the \MCSAT solver and the effect of different explainer
  combinations and propagation. Each curve shows the number of
  benchmarks that the solver variant can solve against the time.}
  \label{fig:timegraph}
\end{figure}

We implemented our approach in the \MCSAT solver within Yices~2~\cite{Dutertre14yices}.
To evaluate its effectiveness, and the impact of the different modules,
we ran the \MCSAT solver with different settings on
the 41,547 QF\_BV benchmarks available in the SMT-LIB
library~\cite{BarST-SMTLIB}.
We used a three-minute timeout per instance.
Each curve in Figure~\ref{fig:timegraph}
shows the number of solved instances for each solver variant;
\textsf{all}: the procedures of Sections~\ref{sec:concatextract} and~\ref{sec:bvarith},
with the bitblasting baseline when these do not apply;
\textsf{bb}:       only the bitblasting baseline;
\textsf{bb+eq}:    procedure of Section~\ref{sec:concatextract} plus the baseline;
\textsf{bb+arith}: procedure of Section~\ref{sec:bvarith} plus the baseline;
\textsf{all-prop}  is the same as \textsf{all} but with no propagation of
bitvector assignments during search.
For reference, we also included the version of the Yices~2 \MCSAT solver
that entered the 2019 SMT
competition\footnote{\url{https://smt-comp.github.io/2019/}}, marked as
\textsf{smtcomp2019}.

% Results: 
% all: 33241
% bb: 31416
% bb+eq: 31930 ()
% bb+arith: 32822
% all-prop: 31818
% Note: these numbers are for a version of the solver (commit d99bbc)
% that is slightly behind the version (commit 4b604f) used to produce the numbers below
% We will update the numbers

The solver combining all explainer modules solved 33260 benchmarks before timeout,
14,174 of which are solved by pure simplification,
and 19,086 of which actually rely on \MCSAT explanations.
14,334 of those are solved without ever calling the default bitblasting baseline
(only the dedicated explainers of Sections~\ref{sec:concatextract} and~\ref{sec:bvarith} are used),
while the other 4,752 instances are solved by a combination of the three explainers.

The results show that
both equality and arithmetic explainers contribute to the effectiveness of the overall solver,
individually and combined.
A bit more than half of the
problem instances involving \MCSAT explanations are fully within the scope of the
two dedicated explainers.
Of course these explainers are still useful beyond that half, in combination with the
bitblasting explainer.
The results also show that the eager \MCSAT value propagation mechanism
introduced in~\cite{Jovanovic:vmcai17} is important for effective solving in practice.

For comparison, we also ran two solvers \CDCLT solvers based on
bitblasting on the same benchmarks and with the same timeout. We
picked Yices~2~\cite{Dutertre14yices} (version 2.6.1)
and Boolector~\cite{NiemetzPWB:CAV18} (version 3.2.0) and we
used the same backend SAT solver for both, namely CaDiCaL~\cite{CaDiCaL}. Yices~2
solved 40,962 instances and Boolector solved 40,763 instances.  We
found 789 instances in the SMTLib benchmarks where our \MCSAT solver
was faster than Boolector by more than 2 sec.  The
\verb=pspace/ndist*= and \verb=pspace/shift1add*= instances are
trivial for \MCSAT (solved in less than 0.25 sec.~each), while
Boolector hit our 3-minute timeout on all \verb=ndist.a.*= instances
and all but 3 \verb=shift1add*= ones.  The \verb=brummayerbiere4=
instances are trivial for \MCSAT (solved in less than 0.03 sec.)
while Boolector ran out of memory in our experimentation (except for one
instance).  Instances with a significant runtime difference in favour
of \MCSAT\ are among \verb=spear/openldap_v2.3.35/*= and
\verb=brummayerbiere/bitrev*= (\MCSAT\ is systematically better),
\verb=float/mult*= (\MCSAT\ is almost systematically better),\linebreak 
\verb=float/div*=, \verb=asp/SchurNumbers/*=,
\verb=20190311-bv-term-small-rw-Noetzli/*=, and \verb=Sage2/*=.
\MCSAT is almost systematically faster on \verb=uclid/catchconv/*= and
faster on more than half of \verb=spear/samba_v3.0.24/*=.

Using an alternative \MCSAT\ approach to bitvector solving,
Zelji\'{c} et al.~reported that
their solver could solve 23704 benchmarks from a larger set of 49971 instances
with a larger timeout of 1200s~\cite{ZeljicWR16}.\footnote{The additional 8424
  benchmarks have since been deleted from the SMT-LIB library as duplicates.}
We have not managed to reproduce the results of Zelji\'{c}'s solver on our Linux server for direct comparison.

To debug the implementation of our explainers,
every conflict explanation that is produced when solving in debug mode
is sent on-the-fly to (non-\MCSAT) Yices~2,
which checks the validity of the clause by bitblasting.
In debug mode, every normalization we perform with the rules of Section~\ref{sec:normalisation}
is also sent to Yices~2 to prove the equality between the original term and the normalized term.
Performance benchmarking was only done after completing, without any red flag, a
full run of \MCSAT in debug mode on the 41,547 QF\_BV benchmarks instances.

% !TEX root = main.tex
% !TEX program = pdflatex

\section{Discussion and Future Work}
\label{sec:conclusion}

The paper presents ongoing work on building an \MCSAT solver for the theory of bitvectors.
We have presented
two main ideas for the treatment of $\BVth$ in \MCSAT,
that go beyond the approach proposed by Zelji\'{c} et al.~\cite{ZeljicWR16}.

First, by relying on BDDs for representing feasible sets, our design
keeps the main search mechanism of \MCSAT generic and leaves
fragment-specific mechanisms to conflict explanation. The explanation
mechanism is selected based on the constraints involved in the
conflict. BDDs are also used to minimize the conflicts, which
increases the chances that a dedicated explanation mechanism can be applied.
BDDs offer a propagation mechanism that differs from those
in~\cite{ZeljicWR16} in that the justification for a propagated
assignment is computed lazily, only when it is needed in conflict
analysis. Computing the conflict core at that point effectively
recovers justification of the propagations.

Second, we propose explanation mechanisms for two fragments of the
theory: the core fragment of $\BVth$ that includes equality,
concatenation and extraction; and a fragment of linear
arithmetic. Compared to previous work on coarsest-base slicing, such
as~\cite{Bruttomesso:ICCAD09}, our work applies the slicing on the
conflict constraints only, rather than the whole problem. This should
in general make the slices coarser, which we expect to positively
impact efficiency. Our work on explaining arithmetic constraints is
novel, notwithstanding the mechanisms studied by
Janota and Wintersteiger~\cite{JanotaW:SMT16} that
partly inspired our Table~\ref{tab:cases}
but addressed a smaller fragment of arithmetic outside of the context of
\MCSAT.

We have implemented the overall approach in the Yices~2 SMT solver.
Experiments show that the overall approach is effective on practical
benchmarks, with all the proposed modules adding to the solver
performance. \MCSAT is not yet competitive with bitblasting, but we are
making progress. The main challenge is devising efficient word-level
explanation mechanisms that can handle all or a least a large fragment
of \BVth. Finding high-level interpolants in \BVth is still an open
problem and our work on \MCSAT shows progress for some fragments of the
bitvector theory. For \MCSAT to truly compete with bitblasting, we will
need interpolation methods that cover larger classes of constraints.

A key step in that direction is to extend the bitvector arithmetic explainer
so that it handles multiplications by constants, then multiplication by evaluable terms,
and, finally, arbitrary multiplications.
Deeper integration of fragment-specific explainers
could potentially help explaining \emph{hybrid} conflicts
that involve constraints from different fragments.
To complement the explainers that we are developing,
we plan to further explore the connection between interpolant generation
and the closely related domain of quantifier elimination,
particularly those techniques
by John and Chakraborty~\cite{John2016} for the bitvector theory.
The techniques by Niemetz et al.~\cite{NiemetzPRBT:CAV18}
for solving quantified bitvector problems
using \emph{invertibility conditions}
could also be useful for interpolant generation in \MCSAT.

% MCSAT principle that
% handling constraints that are unit in a bit-vector variable is in practice
% easier than handling multivariate constraints. This principle probably applies
% less to benchmarks whose constraints feature complex re-uses of a single
% variable, and we anticipate that our approach will apply to software
% verification problem better than to hardware verification problems.

Future work also includes relating our approach to the report by Chihani,
Bobot, and Bardin~\cite{chihani:hal-01531336}, which aims at lifting the CDCL
mechanisms to the word level of bitvector reasoning, and therefore seems very
close to \MCSAT. Finally, we plan to explore integrating our
\MCSAT treatment of bitvectors with other components of SMT-solvers,
whether in the context of \MCSAT or in different architectures. An approach for
this is the recent framework of \emph{Conflict-Driven Satisfiability}
(CDSAT)~\cite{BonacinaGrahamLengrandShankar:17,BonacinaGrahamLengrandShankar:19JAR1},
which precisely aims at
organizing collaboration between generic theory modules.

% Note that on benchmarks that are not entirely in the bit-vector fragment described in this paper,
% our interpolation mechanism needs to collaborate with other interpolation mechanisms.
% Improving the way in which different interpolation mechanisms can collaborate for conflict analysis in MCSAT
% is one of our ongoing research directions.
% The treatment of the \verb=QF_BV/pspace/shift1add.*.smt2= benchmarks is in fact an example of collaboration
% between the interpolation mechanism described in this paper,
% and that which we implemented for the fragment of bit-vector arithmetic consisting of
% extraction, concatenation, and equalities,
% and which we described in~\cite{GrahamLengrandJovanovicSMT17}.

%------------------------------------------------------------------------------
{\small
  \paragraph{Acknowledgments}
  \label{sect:acks}
  The authors thank Aleksandar Zelji{\'c} for fruitful discussions.
  This material is based upon work supported in part by NSF grants 1528153
  and 1816936, and by the Defense Advanced Research Project Agency (DARPA)
  and Space and Naval Warfare Systems Center, Pacific (SSC Pacific) under
  Contract No.~N66001-18-C-4011. Any opinions, findings and conclusions or
  recommendations expressed in this material are those of the author(s)
  and do not necessarily reflect the views of NSF, DARPA, or SSC Pacific.
}

\newpage

\label{sect:bib}

\longv[\bibliographystyle{splncs04}]{\bibliographystyle{good}}

\bibliography{abbrev-short,main,crossrefs}

\longv{
  %% Appendices
  
  \newpage
  \appendix

  \section{Differences with previous workshop presentations}

  The present contribution improves on our previous SMT workshop contributions~\cite{GrahamLengrandJovanovicSMT17,GrahamLengrandJovanovicSMT19}
  as follows:
  \begin{enumerate}
  \item
    Both the concatenation-extraction explainer
    (whose design was described in~\cite{GrahamLengrandJovanovicSMT17})
    and the arithmetic explainer
    (described in~\cite{GrahamLengrandJovanovicSMT19}),
    have seen their scope of application significantly extended
    by the notion of \emph{evaluable term}.
    This can be seen by comparing the fragments' grammars
    with those of~\cite{GrahamLengrandJovanovicSMT17,GrahamLengrandJovanovicSMT19}.
    Evaluable terms can feature any operator of the \BVth\ theory,
    as long as the conflict variable does not appear.
    The implementation (inexistant at the time of~\cite{GrahamLengrandJovanovicSMT17})
    has significant machinery to detect and handle evaluable terms.
  \item
    The arithmetic explainer has been enriched with
    concatenations and upper-bits extractions,
    which were not even broached in~\cite{GrahamLengrandJovanovicSMT19}.
    Regarding extraction, it only addressed lower-bits extraction,
    and even that was not implemented.
    Arbitrary extractions, and concatenations, are entirely new,
    and triggered the design of the algorithm described in Fig.~\ref{fig:Iinverse}.
  \item
    The aggressive normalization applied to conflict cores before they are analyzed,
    presented in Section~\ref{sec:normalisation},
    is also mostly new: only a very limited form was present in~\cite{GrahamLengrandJovanovicSMT19}.
  \item
    Finally, no experimental results were described in~\cite{GrahamLengrandJovanovicSMT17,GrahamLengrandJovanovicSMT19}.
    In fact, no implementation had been developed regarding the design proposed in~\cite{GrahamLengrandJovanovicSMT17}. 
  \end{enumerate}

  \section{Correctness of the concatenation-extraction explainer}
  \input{toappendix}
  \section{Complements on interpolation for bitvector arithmetic}
\label{app:simple}

\subsection{Related work}

Table~\ref{tab:cases} is inspired by Table~1 in Janota and Wintersteiger's SMT'2016 paper~\cite{JanotaW:SMT16}.
We leverage the approach for the purpose of building interpolants,
so in our case the expressions $e_1$, $e_2$, etc are not constants,
but can have variables (with values in model ${\mathcal M}$).
A rather cosmetic difference we make consists in working with intervals that exclude their upper bound,
as this makes the theoretic and implemented treatment of those intervals simpler
and more robust to the degenerate case of bitwidth 1, where $1=-1$.
Another difference is that we take circular intervals,
so that every constraint corresponds to exactly one interval;
as a result,
we do not need the case analyses expressed by the conditions of Table~1 in~\cite{JanotaW:SMT16}.
We do, however, make some new case analyses
to detect when a constraint leads to an empty or full forbidden interval,
since such intervals will be subject to a specific treatment when generating interpolants,
as described in Section~\ref{sec:interp}.

\subsection{Particular case of interpolation with only one bitwidth}

\begin{algorithm}[t]
  \caption{Extracting a covering sequence of intervals}\label{algo:sequence}
  \begin{algorithmic}[1]
    \Function{seq\_extract}{$\{I_1,\ldots,I_m\},{\mathcal M}$}
    \State $\textsf{output}   \gets ()$ \Comment{output initialized with the empty sequence of intervals}
    \State $\textsf{longest}  \gets \Call{longest}{\{I_1,\ldots,I_m\},{\mathcal M}}$ \Comment{longest interval identified}
    \State $\textsf{baseline} \gets \textsf{longest.upper}$ \Comment{where to extend the coverage from}
    \While{$\sem{\mathcal M}{\textsf{baseline}} \not\in \sem{\mathcal M}{\textsf{longest}} $}
    \State $I \gets \Call{furthest\_extend}{\textsf{baseline},\{I_1,\ldots,I_m\},{\mathcal M}}$
    \State $\textsf{output}   \gets \textsf{output},I$ \Comment{adding $I$ to the output sequence}
    \State $\textsf{baseline} \gets I\textsf{.upper}$ \Comment{updating the baseline for the next interval pick}
    \EndWhile
    \If{$\sem{\mathcal M}{\textsf{baseline}} \in \sem{\mathcal M}{\textsf{output.first}}$}
    \State\Return{$\textsf{output}$} \Comment{the circle is closed without the help of $\textsf{longest}$}
    \EndIf
    \State\Return{$\textsf{output},\textsf{longest}$} \Comment{$\textsf{longest}$ is used to close the circle}
    \EndFunction
  \end{algorithmic}
\end{algorithm}
When the intervals $I_1\ldots,I_m$ generated from $C_1,\ldots,C_m$ are all forbidding
values for the same lower-bits extract $\bextract w y$ of the conflict variable $y$,
we know that $\bigcup_{i=1}^m\sem{\mathcal M}{I_i}$ is the full domain $\ZZ/2^w\ZZ$.
%% In the context of MCSAT~\cite{Jovanovic13mcsat},
%% we then backtrack over the construction of ${\mathcal M}$ to try and build a different model;
%% while attempting the new model construction,
%% we will avoid any model satisfying property $E_1\wedge\cdots\wedge E_p$,
%% in order not to fall into a conflict of the same nature as the one we just analysed.
%% With different values for variables $\vec x$ than those specified by ${\mathcal M}$,
%% the intervals $\bigcup_{i=1}^m{I_i}$ could cover the full domain in many different ways.
%% Here, we are trying to symbolically capture the way that ${\mathcal M}$ allows itmakes those intervals a full coverage.
We can then use Algorithm~\ref{algo:sequence}
to extract a sequence $I_{\pi(1)}, \ldots, I_{\pi(q)}$
from $\{I_1\ldots,I_m\}$
(\ie an injective function $\pi$ from $[1;q]$ to $[1;m]$)
that covers $\ZZ/2^w\ZZ$ in the following sense:
$\bigcup_{i=1}^q\sem{\mathcal M}{I_{\pi(i)}}$ is still $\ZZ/2^w\ZZ$
as in model $\mathcal M$
the upper bound of each interval belongs to the next interval in the sequence.
Algorithm~\ref{algo:sequence} relies on the following ingredients:
\begin{itemize}
\item $\Call{longest}{\{I_1,\ldots,I_m\},{\mathcal M}}$ returns an interval among $\{I_1,\ldots,I_m\}$ whose concrete version $\sem{\mathcal M}{I}$ has maximal length;
\item $I\textsf{.upper}$ denotes the upper bound of an interval $I$ (it is \emph{excluded} from $I$);
\item $\Call{furthest\_extend}{a,\{I_1,\ldots,I_m\},{\mathcal M}}$
  returns an interval $I$ among\linebreak $\{I_1,\ldots,I_m\}$ that furthest extends $a$ according to ${\mathcal M}$ (technically,
  an interval $I$ that $\uleq$-maximizes $\sem{\mathcal M}{I\textsf{.upper}-a}$
  among those intervals $I$ such that $\sem{\mathcal M}{a}\in \sem{\mathcal M}{I}$).
\item $\textsf{output.first}$ denotes the first element of a sequence $\textsf{output}$;
\end{itemize}
Algorithm~\ref{algo:sequence} stops with the first interval $I$ that closes the circle,
in that its concrete upper bound $\sem{\mathcal M}{I\textsf{.upper}}$ belongs to $\sem{\mathcal M}{\textsf{longest}}$
(it may or may not close the circle without the help of $\sem{\mathcal M}{\textsf{longest}}$, hence the final \textsf{if...then...else}).
Note that $\bigcup_{i=1}^m\sem{\mathcal M}{I_i}$ is not the full domain if and only if
one of the calls $\Call{furthest\_extend}{a,\{I_1,\ldots,I_m\},{\mathcal M}}$ fails.

\begin{example}\label{ex:3}
  In Example~\ref{ex:2}.2, the coverage algorithm~\ref{algo:sequence} produces the sequence $I_1,I_3,I_2$,
  namely $\interval{x_1}{x_1+1},\ \interval{x_2}{-x_3},\ \interval{-x_3}{x_1-x_3}$,
  since the longest concrete interval is $\sem{\mathcal M}{I_2}$.
\end{example}

\begin{remark}
  The reason why we identify an interval of maximal length is to obtain a \emph{minimal} coverage of the full domain:
  otherwise the last interval added to the sequence could include some of the first ones;
  removing those from the sequence would still produce a covering sequence.\footnote{The issue does not occur in \MCSAT as currently implemented,
    where we have an extra piece of information, namely that the original constraints $C_1,\ldots, C_m$
    form a \emph{core} of the conflict:
    if one of them, say $C_1$, is removed, then $\exists y(C_2\wedge\cdots\wedge C_m)$ evaluates to true in ${\mathcal M}$.
    If one of the intervals, say $I_1$, was not needed for the coverage,
    then $C_1$ would not be in the core.
    Hence in our implementation, $q$ is always $m$ and the sequence is just an ordering of the set of intervals.
    Moreover if one of the intervals is full, then it must be the only interval.
    Still, the algorithm above allows us to produce the ordering.}
  This does not happen when starting the sequence by extending the longest interval,
  but of course there could still be covering sequences with a smaller number of intervals.
\end{remark}

\begin{remark}
  The produced interpolant involves generating constraints $u_i\in I_{i+1}$.
  If $I_{i+1} = \interval{l_{i+1}}{u_{i+1}}$,
  a naive way of expressing $u_i\in I_{i+1}$ would be $(l_{i+1}\uleq u_i\ult u_{i+1})$.
  That would fail to capture the possibility that the intervals overflow.\footnote{A particular case could be made for the interval(s) that overflow(s), expressing the linking property differently,
    but that would actually give a particular role to the constant $0$ in the circular domain $\ZZ/2^w\ZZ$.
    This would weaken the interpolant,
    in the sense that it would rule out fewer models that falsifies $\mathcal A$ ``for the same reason'' ${\mathcal M}$ does.
    Indeed, imagine another model ${\mathcal M}'$ falsifying $\mathcal A$ and leading to concrete intervals 
    $\sem{{\mathcal M}'}{I_1},\ldots,\sem{{\mathcal M}'}{I_m}$
    that only differ from 
    $\sem{{\mathcal M}}{I_1},\ldots,\sem{{\mathcal M}}{I_m}$
    in that all bounds are shifted by a common constant.
    The interpolant that gives a special role to $0$ may not rule out ${\mathcal M}'$,
    whereas the interpolant we produce does.}
\end{remark}

  \section{Example on multiple bitwidths}
\label{ex5}

\begin{example}\label{ex:5}\strut
  Consider a variant of Example~\ref{ex:1}.2 with
  the constraints $C_1, C_2, C_3, C_4$ presented on the first line of Figure~\ref{fig:ex6},
  and model ${\mathcal M}=\{x_1\mapsto 1100,x_2\mapsto 1101,x_3\mapsto 0000\}$.
  The second line is obtained from Table~\ref{tab:cases},
  with the conditions on the third line being satisfied in ${\mathcal M}$.

  \begin{figure}[t]
    \begin{center}
      \begin{tabular}{|c|c|c|c|c|}
        \hline
        \multirow{2}{*}{\textbf{Constraint $C$}}
        & $C_1$
        & $C_2$
        & $C_3$
        & $C_4$\\
        & $\neg(y \eq x_1)$
        & $(x_1 \uleq x_3 + y)$
        & $(\bextract 2 y \uleq \bextract 2 {x_2})$
        & $(\bextract 1 y \eq 0)$\\\hline
        \begin{tabular}c
          {\bf Forbidden}\\
          {\bf interval $I_{C}$}
        \end{tabular}
        & $\interval{x_1}{x_1+1}$
        & $\interval{-x_3}{x_1-x_3}$
        & $\interval{\bextract 2 {x_2}+1}{0}$
        & $\interval{1}{0}$\\\hline
            {\bf Condition ${c}$}
            & $(0\noteq -1)$
            & $(x_1\noteq 0)$
            & $(\bextract2{x_2}\noteq -1)$
            & $(0\noteq -1)$\\\hline
            \begin{tabular}c
              {\bf Concrete}\\
              {\bf interval $\sem{\mathcal M}{I_{C}}$}
            \end{tabular}
            & $\interval{1100}{1101}$
            & $\interval{0000}{1100}$
            & $\interval{10}{00}$
            & $\interval{1}{0}$\\\hline
                {\bf bitwidth $w_i$}
                & \multicolumn2{c|}{$w_1 = 4$}
                & $w_2 = 2$
                & $w_3 = 1$\\\hline
                {\bf Interval layer $\mathcal S_i$}
                & \multicolumn2{c|}{$\mathcal S_1 = \{I_{C_1},I_{C_2}\}$}
                & $\mathcal S_2 = \{I_{C_3}\}$
                & $\mathcal S_3=\{I_{C_4}\}$\\\hline
                \begin{tabular}c
                  {\bf Forbidding}\\
                  {\bf values for}
                \end{tabular}
                & \multicolumn2{c|}{$y$}
                & $\bextract 2 y$
                & $\bextract 1 y$\\\hline
      \end{tabular}
    \end{center}
    \caption{Example with multiple bitwidths}
  \label{fig:ex6}
  \end{figure}

  Algorithm~\ref{algo:sequence2} identifies $I_{C_2}$
  as the longest among $\mathcal S_1$ in model ${\mathcal M}$.
  The next interval among $\mathcal S_1$ covering $(x_1{-}x_3)$ in ${\mathcal M}$ is $I_{C_1}$,
  so $(x_1{-}x_3)\in I_{C_1}$ is added as an interpolant constraint $d_1$.
  Then $x_1{+}1$ is not covered in ${\mathcal M}$ by any interval in $\mathcal S_1$:
  it starts a hole that spans up to $-x_3$.
  The hole $\interval{x_1{+}1}{-x_3}$ has length $0011\ult 2^2$ in ${\mathcal M}$,
  so $(-x_3{-}x_1{-}1\ult 2^2)$ is added as an interpolant constraint $d_2$ and
  a recursive call is made on
  $\mathcal S'_2 = \{I_{C_3},I \}$ and $\mathcal S_3=\{I_{C_4}\}$,
  where $I = \interval{\bextract2{-x_3}}{\bextract2{x_1}{+}1}$.
  The longest interval among $\mathcal S'_2$ in ${\mathcal M}$ is $I_{C_3}$,
  and it upper bound $00$ is covered in ${\mathcal M}$ by $I$,
  so $00\in I$ is added as an interpolant constraint $d_3$.
  Then $\bextract2{x_1}{+}1$ is not covered in ${\mathcal M}$ by any interval in $\mathcal S'_2$:
  it starts a hole that spans up to $\bextract 2 {x_2}+1$.
  The hole $\interval{\bextract2{x_1}{+}1}{\bextract 2 {x_2}{+}1}$ has length $01\ult 2^1$ in ${\mathcal M}$,
  so $(\bextract 2 {x_2}{-}\bextract2{x_1}\ult 2^1)$ is added as an interpolant constraint $d_4$ and
  a recursive call is made on $\mathcal S'_3=\{I_{C_4}, I'\}$
  where $I'=\interval{\bextract 1 {x_2}{+}1}{\bextract1{x_1}{+}1}$.
  Intervals $I_{C_4}$ and $I'$ finally cover $\ZZ/2\ZZ$,
  with $(\bextract1{x_1}{+}1)\in I_{C_4}$ and $0\in I'$ added as interpolant constraints $d_5$ and $d_6$.
  Coming back from the recursive calls,
  $(\bextract 2 {x_2}{+}1)\in I_{C_3}$ and then $-x_3\in I_{C_2}$ are added as interpolant constraints
  $d_7$ and $d_8$.
  The interpolant is
  $\bigwedge_{i=1}^8 d_i \imp\bot$.
\end{example}

}

\end{document}